\documentclass[a4paper,UKenglish,cleveref, autoref, thm-restate]{lipics-v2021}

\title{Adaptive Curves for Optimally Efficient Market Making} 

\author{Viraj {Nadkarni}}{Princeton University, USA \and \url{https://sites.google.com/view/virajnadkarni/home}} {viraj@princeton.edu}{https://orcid.org/0000-0003-2637-3676}{}

\author{Sanjeev Kulkarni}{Princeton University, USA \and \url{https://www.princeton.edu/~kulkarni/} }{kulkarni@princeton.edu}{https://orcid.org/0000-0002-5308-5250}{}

\author{Pramod Viswanath}{Princeton University, USA \and \url{https://ece.princeton.edu/people/pramod-viswanath}} {pramodv@princeton.edu}{https://orcid.org/0000-0003-3171-8667}{}

\authorrunning{V. Nadkarni, S. Kulkarni, and P. Viswanath} 

\Copyright{Viraj Nadkarni, Sanjeev Kulkarni, and Pramod Viswanath} 


\relatedversiondetails[cite={nadkarni2024adaptive}]{Full Version}{https://arxiv.org/pdf/2406.13794} 

\funding{This work was supported by the National Science Foundation via grants CNS-2325477, the Army Research Office via grant W911NF2310147, a grant from C3.AI, a SEAS Innovation grant from Princeton University and a gift from XinFin Private Limited.}

\nolinenumbers 

\EventEditors{Rainer B\"{o}hme and Lucianna Kiffer}
\EventNoEds{2}
\EventLongTitle{6th Conference on Advances in Financial Technologies (AFT 2024)}
\EventShortTitle{AFT 2024}
\EventAcronym{AFT}
\EventYear{2024}
\EventDate{September 23--25, 2024}
\EventLocation{Vienna, Austria}
\EventLogo{}
\SeriesVolume{316}
\ArticleNo{1}

\usepackage{algorithm}
\usepackage{algpseudocode}
\usepackage{graphicx,subcaption}

\usepackage[utf8]{inputenc} 
\usepackage[T1]{fontenc}    
\usepackage{hyperref}       
\usepackage{url}            
\usepackage{booktabs}       
\usepackage{amsfonts}       
\usepackage{nicefrac}       
\usepackage{microtype}      
\usepackage{xcolor}         
\usepackage{amsmath}
\usepackage{amsthm}

\usepackage{prettyref}
\newrefformat{fig}{Figure~[\ref{#1}]}

\newrefformat{app}{\ref{#1}}
\newrefformat{alg}{\ref{#1}}

\begin{document}

\maketitle
\begin{abstract}
Automated Market Makers (AMMs) are essential in Decentralized Finance (DeFi) as they match liquidity supply with demand. They function through liquidity providers (LPs) who deposit assets into liquidity pools. However, the asset trading prices in these pools often trail behind those in more dynamic, centralized exchanges, leading to potential arbitrage losses for LPs. This issue is tackled by adapting market maker bonding curves to trader behavior, based on the classical market microstructure model of Glosten and Milgrom. Our approach ensures a zero-profit condition for the market maker's prices. We derive the differential equation that an optimal adaptive curve should follow to minimize arbitrage losses while remaining competitive. Solutions to this optimality equation are obtained for standard Gaussian and Lognormal price models using Kalman filtering. A key feature of our method is its ability to estimate the external market price without relying on price or loss oracles. We also provide an equivalent differential equation for the implied dynamics of canonical static bonding curves and establish conditions for their optimality. Our algorithms demonstrate robustness to changing market conditions and adversarial perturbations, and we offer an on-chain implementation using Uniswap v4 alongside off-chain AI co-processors.

\end{abstract}

\maketitle

\section{Introduction}
\label{sec:intro}



Market making plays a crucial role in enhancing liquidity in financial systems. In traditional finance, effective market making strategies involve setting buy and sell prices (bid and ask quotes) as narrowly separated as possible, ensuring these quotes closely mirror the asset's true price on a limit order book. This strategy enables market makers to earn a marginal profit. Such efficacy in market making is driven by sophisticated models that analyze trader behavior \cite{GLOSTEN198571,kyleModel,grossmanMiller}. These models have become fundamental in understanding the principles of microeconomics and the microstructure of markets.

\subparagraph*{Market Makers in DeFi} 
In the field of Decentralized Finance (DeFi), the concept of \textit{automated} market making has gained prominence. DeFi employs Automated Market Makers (AMMs), particularly Constant Function Market Makers (CFMMs) \cite{mohanDexPrimer}\cite{angeris2021constant}, offering an alternative to traditional limit order books. This approach reduces the computational effort needed to facilitate trades and ensures liquidity is available for tokens that are less frequently traded. Unlike conventional markets that primarily utilize limit order books for Peer-to-Peer transactions, decentralized markets implement a Peer-to-Pool-to-Peer structure. In this model, Liquidity Providers (LPs) aggregate their resources in a contract, which traders then utilize to meet their liquidity needs. Thus, any DeFi market needs to incentivize both the LPs and the traders to ensure a fair and efficient market is created.

\subparagraph*{Market depth and volatility} 
DeFi markets exhibit a range of distinct characteristics, primarily differentiated by market depth (or liquidity) and price volatility. Notably, the trading volume of stablecoins (approximately \$11.1 trillion) has recently exceeded the transaction volumes of centralized entities like MasterCard and PayPal \cite{stablecoin3}. Markets with significant liquidity, especially those trading stablecoins, are highlighted in this context \cite{univ3}. Such markets typically experience minimal volatility, and their substantial depth minimizes the price impact of retail trades. Conversely, DeFi features hundreds of infrequently traded tokens, which suffer from a lack of liquidity, leading to high volatility and price sensitivity to even small-scale retail trades. These markets are also susceptible to swings in price caused by flash loan transactions \cite{flashLoansAttack}. This paper addresses the optimization of market making strategies for the latter (less liquid) kind of markets.

\subparagraph*{Incentives of LPs} 
A key challenge for Constant Function Market Makers (CFMMs) is motivating Liquidity Providers (LPs) to contribute their tokens to the pool. For this incentive to work, it is crucial for CFMMs to minimize the average losses on pooled assets. Yet, it is widely acknowledged that LPs often incur losses due to fluctuations in reserves \cite{loesch2021impermanent} and a lack of market insights \cite{milionis2022automated}. This paper concentrates on reducing the losses that stem from such informational deficiencies. Specifically, static curves in CFMMs frequently lead to LP losses as a result of arbitrage activities. These losses are intended to be offset by transaction fees, contrasting with centralized exchanges which benefit from higher liquidity and trading volumes but impose lower fees. For example, Binance, a centralized exchange, records a daily trading volume of approximately \$15 billion, significantly higher than Uniswap’s \$1.1 billion \cite{tradeVolComparisons}, the largest decentralized exchange. The lower liquidity on platforms like Uniswap results in less current prices, making them more susceptible to arbitrage losses.

 \subparagraph*{Arbitrage Loss} 
 The specific type of arbitrage loss known as loss-versus-rebalancing (LVR) can be quantified in certain scenarios \cite{milionis2022automated}, and these losses continue to occur despite the implementation of trading fees \cite{milionis2023automated}. In the case of a generic market maker who sets bid and ask prices for a volatile asset, arbitrage losses are defined relative to the asset's true market price. An arbitrageur engages in a buy transaction when the market price surpasses the ask price, and in a sell transaction when it drops below the bid price. The resultant loss for the market maker is calculated as the product of the price difference and the volume of the asset traded.

\subparagraph*{Trader behavior} 
In traditional financial systems, arbitrage-related losses are conceptualized as \textit{adverse selection} costs, which arise from interactions with \textit{informed traders}—those who are privy to the external market price, akin to arbitrageurs. A market maker achieves optimal operation by balancing these costs against the profits gained from \textit{uninformed traders}, also known as \textit{noise traders}. This balancing principle was first delineated by Glosten and Milgrom \cite{GLOSTEN198571}. Within the Decentralized Finance (DeFi) ecosystem, trading parties are differentiated into \textit{toxic} and \textit{non-toxic} order flows, which correspond to informed and uninformed traders, respectively \cite{nezlobinToxicity,crocToxicity}. We extend this model to a more nuanced framework where traders are categorized along a continuous spectrum of information awareness, ranging from highly informed to completely uninformed, rather than being strictly classified as either \textit{toxic} or \textit{non-toxic} (\prettyref{sec:model}).

\subparagraph*{CFMMs as prediction markets} 
For CFMMs, a significant portion of their losses also arises from the need to encourage traders to disclose their genuine price perceptions during transactions. In essence, CFMMs provide compensation to informed traders in exchange for their crucial market insights, which mirrors the principles of market scoring rules utilized in prediction markets to extract valuable information \cite{frongillo2023axiomatic}.

\subparagraph*{Conditions for optimality} 
A straightforward approach to reduce losses to arbitrage would be aligning the marginal price exactly with the external market price, which would require real-time data from a \textit{price oracle} \cite{dodoOracle}. Yet, integrating oracles with market making strategies can lead to potential frontrunning risks \cite{frontrunningOracles} and necessitates reliance on centralized, potentially manipulable external entities \cite{Eskandari_2021,oraclesProblem}. To circumvent these issues, our framework explicitly excludes the use of oracles. The objective is to deduce the hidden market price by analyzing trade history data, aiming for maximum efficiency in terms of data utilization. Further, the market maker uses this to adaptively set its bonding curve so that the loss to arbitrageurs is as close to zero as possible, ensuring an optimally efficient market. The market maker turning a profit would be undesirable since this would allow a competitor to undercut its prices and take away their order flow. In other words, it should quote an efficient and competitive market price, given only the information it has in form of the trading history. Keeping this objective in mind, we outline the key contributions of this work.

\subsection{Our contributions}

\subparagraph*{Optimal algorithms for adapting curves} (\prettyref{sec:variable_trade}) We provide the differential equation that the demand curve of an optimally efficient market should follow (\prettyref{thm:variab}). When the statistics governing trader and price behavior are known and Gaussian/Lognormal, we show that this differential equation can be solved exactly using a dynamic bonding curve that changes its operating point using the Kalman Filter (\prettyref{thm:2} and \prettyref{thm:3}). 

\subparagraph*{Adapting to unknown market conditions} When the statistics governing trader and price behavior are unknown, we extend the previous approach by using an Adaptive Kalman Filter. We empirically show that both these approaches suffer significantly lower arbitrage losses compared to a static CFMM. (\prettyref{sec:akf})

\subparagraph*{Robustness to adversarial manipulation} In presence of irrational traders that seek to make the price of the market maker deviate from the external market, we present a robust version of the adaptive curve algorithm that tolerates upto 50\% of trader population being adversarial. (\prettyref{sec:adverse})

\subparagraph*{Comparisons with static curves} (\prettyref{sec:static_implications}) We provide theoretical comparisons of static and the proposed adaptive curve models by showing that the error in the adaptive AMM price, when viewed as an oracle, decays with more trades, while that of a static curve remains unchanged (\prettyref{thm:static_comparison}). 

\subparagraph*{Implied dynamics of static curves} (\prettyref{sec:static_implications})We derive a differential equation \prettyref{eq:inverse1} that governs the \textit{implied} dynamical model given the static CFMM curves that are used in practice. We show that these CFMMs can only be optimally efficient in a model where inter-block time for the underlying blockhchain vanishes, the CFMM has more liquidity than the external market, and that the trader and price behaviour is severely constrained. (\prettyref{thm:static_implied}) 

\subparagraph*{On-chain Implementation} We specify the end-to-end system design for our proposed market maker. Furthermore, we provide an implementation of our algorithm using the recently released Uniswap v4 \cite{uniswapv4} platform and an off-chain machine learning co-processor Axiom \cite{axiom}. This co-processor provides guarantees that the algorithms derived in this work are executed, and the result is put securely on-chain. (\prettyref{sec:implementation})

\section{Related work}

In this section, we reprise relevant literature surrounding the problem formulated in this paper. Although the motivation of the problem stems from literature studying AMMs in DeFi, our formulation derives heavily from classical works in market microstructure. The algorithms that we present in our work derive heavily from literature on control and robust filtering.

\subparagraph*{Automated Market Makers: }Automated Market Makers (AMMs), particularly in the form of Constant Function Market Makers (CFMMs) \cite{sokDex, mohanDexPrimer}, are designed to incentivize trades that align prices with a more liquid external market \cite{Angeris_2020}. However, this mechanism imposes costs on CFMM liquidity providers while generating profits for arbitrageurs \cite{evans2021optimal, Heimbach_2022, tangri2023generalizing, cartea23predictable, cartea24strategic}. This arbitrage profit, often quantified as "loss-versus-rebalancing" in scenarios where only arbitrageurs (informed traders) interact with the market maker, is proportional to external price volatility \cite{milionis2022automated}. Various methods have been proposed to capture this loss, including on-chain auctions \cite{mcmenamin2022diamonds} and the application of auction theory to dynamically recommend ask and bid prices for an AMM \cite{milionis2023myersonian}. Another recent study \cite{goyal2023finding} suggests an optimal curve for a CFMM based on liquidity providers' price beliefs. However, this study does not consider a dynamic model where traders react to market maker price settings. In \cite{milionis2023myersonian}, a dynamic trading model is examined, deriving optimal ask and bid prices. Yet, this approach necessitates the market maker's knowledge of underlying model parameters, limiting adaptability to market conditions. Additionally, while \cite{milionis2023myersonian} focuses on a monopolistic market maker, our work examines a \textit{competitive} market maker. Our research aligns closely with \cite{nadkarni2023zeroswap}, which addresses competitive market conditions but only within a liquid, non-volatile market with restricted trade sizes. Reinforcement learning algorithms to adapt CFMM bonding curves have been explored in \cite{churiwala2022qlammp}, though the primary objectives there are fee revenue control and minimization of failed trades.

\subparagraph*{Optimal market making: }The trader behavior model we employ is derived from the Glosten-Milgrom model \cite{GLOSTEN198571}, which is widely used in market microstructure literature. However, we modify it to incorporate a continuously changing external price. Several subsequent studies \cite{das2005learning, das2008adapting} develop optimal market-making rules within a modified Glosten-Milgrom framework, but they assume that the underlying model parameters are known and that external price changes are communicated to the market maker. A more data-driven reinforcement learning approach is taken in \cite{chan2001}, but their reward function presupposes direct information about the external hidden price, whereas we assume no access to a price oracle. Another aspect of optimal market making in traditional market microstructure literature focuses on inventory management \cite{hoAndStoll, avellaneda2008}, rather than the information asymmetry between traders and market makers. Our goal is to design a market maker that mitigates losses due to information asymmetry, similar to the Glosten-Milgrom model, without imposing inventory constraints. The Glosten-Milgrom model has been considered for AMMs in DeFi, though only for individual trades \cite{aoyagi2020LP, angeris2020does}.

\subparagraph*{Optimal filtering and control: } The classical filtering and control literature underpins many contemporary automated systems, offering theoretical foundations and practical applications for dynamic system regulation. Kalman filtering, a robust statistical method, is widely used for estimating the state of a linear dynamic system from incomplete and noisy measurements \cite{kalman}. This technique combines real-time measurements with prior estimates to produce updated predictions, proving crucial for systems requiring high accuracy and responsiveness, such as navigation systems, aerospace engineering, and automated trading. In control theory, concepts like optimal control and feedback mechanisms are fundamental for designing systems that maintain desired output levels despite external condition changes. These principles have been extensively explored in works such as \cite{wiener1948cybernetics}, and further developed through modern control theories addressing non-linearities and uncertainties in system dynamics \cite{zhou1996robust}. The integration of filtering and control methodologies has led to the development of Adaptive Kalman Filtering \cite{mehra1970adaptive}, which adjusts its parameters based on observed errors, enhancing performance in varying conditions. Adaptive Kalman Filtering has seen applications in diverse areas, including robotics \cite{thrun2005probabilistic}, automotive systems \cite{gustafsson2002particle}, and finance \cite{elliott2006forecasting}, illustrating its versatility and robustness in handling dynamic, uncertain environments.

\section{Preliminaries and model}
\label{sec:model}

We now describe the framework used for modeling trader behavior in response to the evolution of an external price process, that is hidden from the market maker, and the prices set by the market maker. We also state the objective that the market maker seeks to optimize, and provide the motivation behind it. The model and the objective are based on the canonical Glosten-Milgrom model \cite{GLOSTEN198571} studied extensively in market microstructure literature. We assume that the market maker has access to an inventory of the asset and numeraire. The price of the asset is expressed in terms of the numeraire. We assume that time is discrete and indexed by $t$.

\subparagraph*{External price process:} The external price process $p_{ext}^t$ of a risky asset is assumed to follow a discrete time random walk, where the distribution of a price jump at any $t$ is parametrized by $\sigma$. That is, we have 
\begin{align}
    p_{ext}^{t+1} &= p_{ext}^{t} + \Delta p_{ext}^{t}
\end{align}
where $\Delta p_{ext}^{t}$ is i.i.d. $\sim \mathcal{D}_{\sigma,p_{ext}^{t}}$. Intuitively, the parameter $\sigma$ can be thought of as a measure of volatility of the external price. Further, the random process $p_{ext}^{t}$ can either represent the price of the asset in a larger and much more liquid exchange, or some underlying ``true'' value of the asset. In both cases, we assume that it is hidden from the market maker. To put it in DeFi terms, the market maker does not have an access to any ``price oracle'' that can tell it information about this external price.

\subparagraph*{Market Maker:} The market maker consists of a pool containing an asset and a numeraire. It then publishes a {demand curve} $g_t(p)$ \cite{milionis2023myersonian} before each trade. All traders have full access to the demand curve at any time. The demand curve specifies how the price of the asset changes based on the inventory available. In particular, $g(p_0^t)$ specifies the amount of asset when the initial operating point of the demand curve is $p_0^t$. That is, the price of an infinitesimal amount $dx$ of asset is $p_0^t dx$. The amount of numeraire in the inventory at the same operating point is $-\int_{0}^{p_0^t} p dg(p)$. For an incentive compatible market maker, we constrain $g(.)$ to be non-increasing (Proposition 2.1 in \cite{milionis2023myersonian}). Equivalently, the market maker may be represented by the canonical \textit{bonding curve} $\phi_t(x,y)$ \cite{sokDex} which is a function (of reserves $x,y$ of the asset, numeraire respectively) that stays constant over any single  trade.


\subparagraph*{Trader behavior: }We assume that a trader appears at every time step $t$, and sees a noisy version of the external price in each time step, denoted by $p_{trad}^t$, with the noise distribution being parametrized by $\eta$.
\begin{align}
    p_{trad}^t &= p_{ext}^t + \Delta p_{trad}^{t}
\end{align}
where $\Delta p_{trad}^{t}$ is i.i.d. noise $\sim \mathcal{D}_{\eta,p_{ext}^{t}}$ sampled at each time step. Intuitively, the parameter $\eta$ can be construed as measuring the level of ``toxicity'' or informed nature of the traders.

\subparagraph*{Trade actions:} As mentioned before, the market maker publishes a demand curve $g_t(p)$ at each time step. The trader performs a trade that brings the operating point of the demand curve from $p_0^t$ to $p_{trad}^t$. Thus, the trader behaves as if it is performing arbitrage between a market with price $p_{trad}^t$ and the demand curve published. For instance, if the operating point $p_0^t$ of the market maker is less than the trader price $p_{trad}^t$, then a trader would buy asset from the market maker to sell on the external market until the operating point shifts to $p_{trad}^t$. 

\subparagraph*{Objective:} Our objective is to design an algorithm to set ask and bid prices for the market maker, such that the expected loss with respect to the external market is minimized and the market maker stays competitive. Taking inspiration from \cite{GLOSTEN198571}, but extending to the modified trader behaviour, we get the following condition
\begin{align}
    p^t &= E[p_{ext}^t|\mathcal{H}_{t-1}, p^t_{trad}]\ \ \ \ \forall p^t_{trad} \label{eq:gm_price_modified}
\end{align}
where $\mathcal{H}_{t-1} = \langle (p_{trad}^{\tau},g_{\tau}(.))\rangle_{\tau=0}^{t-1}$ is the history of trades and demand curves until time $t-1$, and $p^t$ is the net price of the trade (the ratio of change in the reserves of the numeraire with the change in the reserves of the asset).

\subparagraph*{Interpreting the objective: }Setting the prices as per the above objective makes the expected loss of the market makers to traders vanish, since
\begin{align}
    E[(p_{ext}^t-p^t)| \mathcal{H}_{t-1}] = 0\label{eq:monetary_loss}
\end{align}
Note that the expression above quantifies the expected loss for per unit trade of the asset.

The market maker can obtain a strictly positive profit by choosing a curve with slightly higher prices (than the ones obtained in \prettyref{eq:gm_price_modified}) on the ask side or slightly lower prices on the bid side (in other words, by choosing a CFMM with a higher curvature \cite{angeris2020does}). However, doing this would make it less competitive, since any other market maker with slightly greater bid or a slightly lesser ask would offer a better price and take away the trade volume. Although we do not explicitly model other market makers, their presence is implicit in setting prices according to \prettyref{eq:gm_price_modified}. This equation represents the ideal conditions for capital efficiency, where both the trader gets the best price possible while the market maker avoids a loss.

Also, note that the market maker incurs a loss in \textit{every trade} made by an perfectly informed trader. Thus, to make the expected loss vanish, it should learn to set prices so that the loss to more informed traders is balanced by the profit obtained from less informed traders. The equation \prettyref{eq:gm_price_modified} can also be interpreted as striking this balance.


\color{black}

\section{Differential Equation for the optimal curve}
\label{sec:variable_trade}

  As mentioned in \prettyref{sec:model}, we make use of the general \textit{demand curve} formulation for market makers. Assume that the amount of asset in reserves at any price $p_0^t$ is $g_t(p_0^t)$, while the amount of numeraire is $-\int_0^{p_0^t} p dg_t(p)$, where the function $g_t$ needs to be non-increasing for the AMM to be incentive compatible \cite{milionis2023myersonian}. Here, $p_0^t$ is the initial operating point of the AMM at the beginning of the time slot $t$. Let $f_t(p)$ be the belief (probability distribution function) of the AMM over the price $p$ of the asset at time $t$. Let $f_\eta(p)$ be the distribution of noise through which the price is observed by the traders. Let $p_{trad}^t$ be the price that the trader observes. Then, the trader would make a trade with the AMM such that the marginal price of the asset just after the trade is $p_{trad}^t$. We now solve the optimal market conditions given by \prettyref{eq:gm_price_modified}.

In this case, the effective price of a trade at time $t$ is given by
\begin{align}
    p^t &= \frac{-\int_{p_0^t}^{p_{trad}^t} p dg_t(p)}{g_t(p_0^t)-g_t(p_{trad}^t)},\label{eq:gm_lhs}
\end{align}
where $p_0^t$ is the operating point of the AMM just before the trade. Further, by definition of conditional expectation of the external market price, we also have
\begin{align}
    E[p_{ext}^t|\mathcal{H}_{t-1}, p^t_{trad}] &= \frac{\int_0^\infty pf_{\eta}(p_{trad}^t-p)f_t(p)dp}{\int_0^\infty f_{\eta}(p_{trad}^t-p)f_t(p)dp}. \label{eq:gm_rhs}
\end{align}
since the noise at time $t$ is independent of the external price at time $t$.
Substituting \prettyref{eq:gm_rhs} and \prettyref{eq:gm_lhs} in \prettyref{eq:gm_price_modified} gives us
\begin{align}
    \frac{-\int_{p_0^t}^{p_{trad}^t} p dg_t(p)}{g_t(p_0^t)-g_t(p_{trad}^t)} &= \frac{\int_0^\infty pf_{\eta}(p_{trad}^t-p)f_t(p)dp}{\int_0^\infty f_{\eta}(p_{trad}^t-p)f_t(p)dp} \ \ \ \ \forall p^t_{trad} \label{eq:imp_gm1}
\end{align}
Rearranging gives us 
\begin{align}
    g_t(p_{trad}^t) = g_t(p_0^t) + \frac{\int_{p_0^t}^{p_{trad}^t}pg'_t(p)dp}{\beta(p_{trad}^t)}\ \ \ \ \forall p^t_{trad} \label{eq:demand_gm1}
\end{align}
where $\beta_t(p_{trad}^t) = \frac{\int_0^\infty pf_{\eta}(p_{trad}^t-p)f_t(p)dp}{\int_0^\infty f_{\eta}(p_{trad}^t-p)f_t(p)dp} $. On the other hand, if we let $p_{trad}^t \rightarrow p_0^t$ in \prettyref{eq:imp_gm1}, we get
\begin{align}
    p_0^t &= \beta_t(p_0^t)\label{eq:initial_cond1}
\end{align}
Thus, using \prettyref{eq:demand_gm1} and \prettyref{eq:initial_cond1}, the demand curve of the optimal market maker can be given by the following theorem.

\begin{theorem}\label{thm:variab}
Let the amounts of asset and numeraire in the reserves of a market maker be $x_0^t,y_0^t$. Then, the optimal demand curve $g_t(p)$ for a market maker obeys the following differential equation
\begin{align}
    (\beta_t(p)-p)g_t'(p) + \beta_t'(p) g_t(p) - \beta_t'(p) x_0^t &= 0 \label{eq:main1}
\end{align}
with the constraints $\lim_{\delta\rightarrow 0^-}g(p_0^t+\delta) \geq x_0^t$ and $\lim_{\delta\rightarrow 0^-}-\int^{p_0^t+\delta}_0 p dg(p) \geq y_0^t$. This equation can be solved separately for $p>p_0^t$ and $p < p_0^t$.
Furthermore, the initial operating point $p_0^t$ is a solution to the fixed point equation
\begin{align}
    p &=\beta_t(p). \label{eq:main2}
\end{align}
\end{theorem}

The differential equation \prettyref{eq:main1} is obtained by simply applying the Leibniz rule of the derivative of a definite integral on \prettyref{eq:demand_gm1}. Note that we can rewrite \prettyref{eq:demand_gm1} as
\begin{align}
    \int_{p_0^t}^{p_{trad}^t} \left(1-\frac{p}{\beta_t(p_{trad}^t)}\right)dg_t(p) = 0\label{eq:inteq}
\end{align}
We see that a discontinuity at $p_0^t$ implies that $dg_t(p)$ is not a differential but is a negative real number. Since $p=\beta_t(p)$ at $p_0^t$, we have the term inside the integral $\rightarrow 0$ as $p_{trad}^t \rightarrow p_0^t$.
This implies that a discontinuity in $g_t(p)$ is allowed at $p_0^t$. Further, since the \prettyref{eq:inteq} holds for $p_{trad}^t > p_0^t$ and for $p_{trad}^t < p_0^t$, the differential equation can be solved separately for $p>p_0^t$ and $p < p_0^t$. 

Finally, assuming that the AMM finds solutions to \prettyref{eq:main1} and \prettyref{eq:main2}, the trader is free to move the market to $p_{trad}^t$. The market maker then simply uses Bayes' rule to update its beliefs over the external prices as 
\begin{align}
    f_{t+1}(p) = \frac{f_{\eta}(p_{trad}^t-p)f_t(p)}{\int_0^\infty f_{\eta}(p_{trad}^t-p)f_t(p)dp} \label{eq:belief_update}
\end{align}
 The updated belief $ f_{t+1}(p)$ is now used to compute $\beta_{t+1}(p)$, thus completing the market making algorithm.  Equations \prettyref{eq:main1}, \prettyref{eq:main2} and \prettyref{eq:belief_update} describe the complete dynamics of an optimally efficient market maker.

\section{Optimal Solutions for special cases} 

In this section, we present solutions to the differential equation derived in \prettyref{thm:variab} for some special cases: Gaussian and Lognormal price jumps and trader noises. That is, we assume that both $\Delta p_{ext}^{t}$ and $\Delta p_{trad}^{t}$ follow a Gaussian or Lognormal distribution. Under these assumptions, we get the following key results.

\subsection{Kalman Filter algorithm for known market parameters}\label{sec:kf}

\begin{theorem}\label{thm:2}
    If $\Delta p_{ext}^{t} \sim \mathcal{N}(0,\sigma^2)$ and $\Delta p_{trad}^{t} \sim \mathcal{N}(0,\eta^2)$ where $\sigma,\eta$ are known to the market maker, then the fixed point equation $p_0^t = \beta_t(p_0^t)$ \prettyref{eq:main2} has a unique solution given by the Kalman filter \cite{kalman} estimate of $p_{ext}^t$, that is, we have
    \begin{align}
        p_0^t &= E[p_{ext}^{t}|\mathcal{H}_{t-1}].
    \end{align}
    Further, the differential equation \prettyref{eq:main1} has a family of solutions, given by
    \begin{align}
        g_t(p) = \begin{cases}
        x_0^t + \frac{y_0^t}{p_0^t} \ \ &\mathrm{if\ }p\leq p_0^t\\
         \max(0,\Tilde{x}_0^t - C_t (p-p_0^t)^{\frac{K_t}{1-K_t}}) \ \ &\mathrm{if\ }p> p_0^t
    \end{cases}\label{eq:soln1}
    \end{align}
    where $K_t$ is the Kalman gain, and $C_t, \Tilde{x}_0^t$ are non-negative constants, such that $\Tilde{x}_0^t\leq x_0^t$.
\end{theorem}

The above solution corresponds to a CFMM with the initial slope of the bonding curve given by the Kalman estimate of $p_{ext}^t$ by treating $p_{trad}^t$ as noisy observations.  
Thus, we only need to calculate a single quantity $E[p_{ext}^t|\mathcal{H}_t]$ (see Algorithm \prettyref{alg:kf}), and set the other parameters of the demand curve according to \prettyref{eq:soln1}. We get a similar solution if we assume that the external price follows geometric Brownian motion.

\begin{theorem}\label{thm:3}
    If $\log \frac{p_{ext}^{t}}{p_{ext}^{t-1}} \sim \mathcal{N}(0,\sigma^2)$ and $\log \frac{p_{trad}^{t}}{p_{ext}^{t}} \sim \mathcal{N}(0,\eta^2)$ where $\sigma,\eta$ are known to the market maker, then the fixed point equation $p_0^t = \beta_t(p_0^t)$ \prettyref{eq:main2} has a unique solution which is a function of the Kalman filter \cite{kalman} estimate of $\log p_{ext}^t$, that is, we have
    \begin{align}
        p_0^t &= \exp\left(E[\log p_{ext}^{t}|\mathcal{H}_{t-1}]+ \frac{P_{t|t}}{2(1-K_t)}\right),
    \end{align}
    where $K_t$ is the Kalman gain, and $P_{t|t}$ is the variance of the Kalman estimate of $\log p_{ext}^t$.
    Further, the differential equation \prettyref{eq:main1} has a solution given by
    \begin{align}
        g_t(p) = \begin{cases}
        x_0^t + \frac{y_0^t}{p_0^t} \ \ &\mathrm{if\ }p\leq p_0^t\\
         \max(0,\Tilde{x}_0^t - C_t (p^{1-K_t} - \kappa_t)^{\frac{K_t}{1-K_t}}p^{K_t}) \ \ &\mathrm{if\ }p> p_0^t
    \end{cases}\label{eq:soln2}
    \end{align}
     where $C_t,\kappa_t,\Tilde{x}_0^t$ are non-negative constants such that $\Tilde{x}_0^t\leq x_0^t$.
\end{theorem}

\begin{algorithm}[t]
\caption{A Kalman filter based algorithm to adapt the AMM curve}\label{alg:kf}
\begin{algorithmic}[1]
\Require Known $\eta,\sigma$, Reserves $x_0,y_0$
\State $t \gets 0$
\State $T \gets$ Number of total time slots
\State Initial price estimate $p^{0|0}_{ext} \gets p_{ext}^0$
\State $P_{0|0} = 0$
\While{$t \leq T$}
\State $\theta_t \gets \frac{p^{t-1|t-1}_{ext} x_{t-1}}{p^{t-1|t-1}_{ext} x_{t-1} + y_{t-1}}$
\State Publish bonding curve $x^{\theta_t} y^{1-\theta_t} = x_{t-1}^{\theta_t}y_{t-1}^{1-{\theta_t}}$
\State Observe trader action $p_{trad}^t$
\State $K_t \gets \frac{P_{t-1|t-1} + \sigma^2}{P_{t-1|t-1} + \sigma^2 + \eta^2}$ \Comment{Update Kalman gain}
\State $p_{ext}^{t|t} \gets (1-K_t)p_{ext}^{t-1|t-1} + K_t p_{trad}^t$\Comment{Update Kalman estimate of the external price}
\State $P_{t|t} = (1-K_t)(P_{t-1|t-1}+\sigma^2)$\label{line1}\Comment{Update Kalman uncertainty}
\EndWhile
\end{algorithmic}
\end{algorithm}
\subparagraph*{Family of optimal demand curves: }We note that both \prettyref{thm:2} and \prettyref{thm:3} recommend a \textit{family} of demand curves that satisfy the Glosten-Milgrom condition \prettyref{eq:gm_price_modified}. The simplest curve in this family is the one where $\Tilde{x}_0^t = C_t = 0$. This gives the simple demand curve which is constant except for a discontinuity at $p_0^t$. This corresponds to a constant sum market maker with bonding curve $y + p_0^t x = k$ with an adaptive slope given by the Kalman estimate of the external price conditioned on trader behavior.

\subparagraph*{An apparent contradiction, and a resolution:} A feature of the optimal demand curves derived is the fact that they do not make it possible for a trader to express any price $p_{trad}^t$ between $0$ and $\infty$. For instance, if we take the constant sum instance of the family of curves, it only allows the trader to express if the price of the trade is greater than or less than $p_0^t$, but not the exact value of $p_{trad}^t$, which contradicts our assumption that the market maker can observe all of the history of trader prices $\langle p_{trad}^\tau\rangle_{\tau=1}^t$. We can resolve this contradiction by approximating the optimal demand curve as a sum of two demand curves $g_t(p)= g^{opt}_t(p)+g^{exp}_t(p)$, where the former curve is the \textit{optimal} solution with most of the liquidity, and the latter curve with low liquidity to improve its price \textit{expressiveness}. For instance, we can have $g_t(p)= (1-\epsilon)g^{CSMM}_t(p)+ \epsilon g^{CPMM}_t(p)$ where we have
\begin{align}
        g_t^{CSMM}(p) &= \begin{cases}
        x_0^t + \frac{y_0^t}{p_0^t} \ \ &\mathrm{if\ }p\leq p_0^t\\
         0 \ \ &\mathrm{if\ }p> p_0^t
    \end{cases}\\
    g^{CPMM}_t(p) &= 1/\sqrt{p} &
\end{align}
with $\epsilon <<1$. This combines the demand curves of the constant sum market maker (the optimal solution) and a constant product market maker (the expressive solution) as an approximation.

\subparagraph*{Other practical approximations:} Another way we can parametrize the curve of the market maker is to choose it from a family of curves such that the initial marginal price matches $p_0^t$ as prescribed by \prettyref{thm:2} and \prettyref{thm:3}, and processing a trade on any curve in that family is computationally simple. To that end, we can use the Constant Mean Market Makers \cite{evans2020liquidity, balancerWP} with its weighting factor as our variable parameter. Note that the CMMMs performs trades along the curve $x^{\theta}y^{1-\theta} = k$ where $k$ is a constant, and $x,y$ are the quantities of the asset and the numeraire in the market maker \cite{evans2020liquidity}. This ensures that no trade can exhaust either the asset or the numeraire from the AMM reserves. We know that the marginal price of the asset at any state of the reserves is given by $p=\frac{\theta y}{(1-\theta) x}$. Therefore, in our case, we set the value of parameter as $\theta_t = \frac{p_0^t x}{p_0^t x + y}$. This ensures that the starting price of any trade is $p_0^t$, and the market maker can only get a better price than that for a large trade.

\subsection{Adaptive Kalman Filter algorithm for unknown market parameters}\label{sec:akf}
\subparagraph*{Need for more adaptivity: }A major assumption while solving for the optimal demand curve in the Gaussian/Lognormal model was that the market parameters $\sigma,\eta$ that control the variances of the price jump and noise were known to the AMM. These can indeed be obtained by analysing historical trading data in any market, and can be assumed to change slowly on the timescale that prices undergo changes. However, this assumption might not always hold for assets or tokens that are less well known or have no historical data. To deal with this case, we propose a modification to Algorithm \prettyref{alg:kf}. This ensures that we simultaneously estimate the parameters $\eta,\sigma$ and hence help estimate the hidden external market price. To that end, we observe that we can write the likelihood function of all random variables and parameters of our model at time $t$ as follows
\begin{align}
    L_t = L(\langle p_{trad}^\tau,p_{ext}^\tau\rangle_{\tau=1}^t,\eta,\sigma) &= \prod_{\tau=1}^t \frac{1}{2\pi\sqrt{\eta\sigma}}\exp{\left(-\frac{(p_{ext}^\tau-p_{ext}^{\tau-1})^2}{2\sigma^2}-\frac{(p_{trad}^\tau-p_{ext}^{\tau})^2}{2\eta^2}\right)},
\end{align}
which gives us the conditional log-likelihood given the trader actions as
\begin{align}
    E[\log L_t | \langle p_{trad}^\tau\rangle_{\tau=1}^t] &= -\frac{t\log\sigma}{2} -\frac{t\log\eta}{2} - \frac{\sum_{\tau=1}^t A_\tau}{2\sigma^2} - \frac{\sum_{\tau=1}^t B_\tau}{2\eta^2},\label{eq:log_like1}
\end{align}
where $A_\tau,B_\tau$ are given by 
\begin{align}
    A_\tau &=  E[(p_{ext}^\tau)^2|\langle p_{trad}^\tau\rangle_{\tau=1}^t] + E[(p_{ext}^{\tau-1})^2|\langle p_{trad}^\tau\rangle_{\tau=1}^t] + 2E[p_{ext}^\tau p_{ext}^{\tau-1}|\langle p_{trad}^\tau\rangle_{\tau=1}^t]\label{eq:A},\\
    B_\tau &= (p_{trad}^\tau)^2 + E[(p_{ext}^{\tau})^2|\langle p_{trad}^\tau\rangle_{\tau=1}^t] + 2p_{trad}^\tau E[p_{ext}^\tau|\langle p_{trad}^\tau\rangle_{\tau=1}^t]\label{eq:B}.
\end{align}
\subparagraph*{Estimating the unknowns: }We now estimate the market parameters $\sigma,\eta$ using the EM algorithm \cite{emog79}. This can be done by first setting the terms $A_\tau, B_\tau$ using \prettyref{eq:A} and \prettyref{eq:B} with the expectations on the RHS calculated via forward and backward runs of the Kalman filter assuming an initial guess estimate of $\sigma,\eta$. The forward runs of the algorithm involve computing $E[p_{ext}^\tau|\langle p_{trad}^\tau\rangle_{i=1}^\tau]$ for all $\tau = 1, \cdots, t$. These are estimates of the external price given only the data in the past. This can be done using the Kalman filter updates given in Algorithm \prettyref{alg:kf}. Next, we use the Rauch-Tung-Striebel smoother \cite{shumway1982approach}, an essentially backward run of the Kalman filter algorithm given all the statistics obtained from the forward run. This computes statistics such as $E[p_{ext}^\tau|\langle p_{trad}^\tau\rangle_{i=1}^t]$, that is, the estimate of the external price in the past given \textit{all} of the observations till the present time slot. This evaluates all of the terms in $A_\tau$ and $B_\tau$, and completes the E-step of the EM algorithm.


After that, we use \prettyref{eq:log_like1} to find values of $\sigma,\eta$ that maximize the conditional log-likelihood function. This involves setting the gradient of the expected log likelihood function to zero 
\begin{align}
    \nabla_\sigma E[\log L_t | \langle p_{trad}^\tau\rangle_{\tau=1}^t] = 0,\
     \nabla_\eta E[\log L_t | \langle p_{trad}^\tau\rangle_{\tau=1}^t] = 0.
\end{align}
While doing this, we ignore the dependence of $A_\tau,B_\tau$ on $\eta,\sigma$ to obtain
\begin{align}
     \sigma^* = \sqrt{2\frac{\sum_{\tau=1}^t A_\tau}{t}},\ \eta^* = \sqrt{2\frac{\sum_{\tau=1}^t B_\tau}{t}}
\end{align}
which completely specifies the M-step of the EM algorithm.
This has been summarized in Algorithm \prettyref{alg:akf}.

\begin{algorithm}[t]
\caption{Adaptive Kalman filter : EM algorithm estimating unknown parameters at time step $t$}\label{alg:akf}
\begin{algorithmic}[1]
\Require Trader data $\langle p_{trad}^\tau\rangle_{\tau=1}^t$, Error Tolerance $\epsilon$
\State $\sigma \gets $ Initial guess $\sigma_0$
\State $\eta \gets $ Initial guess $\eta_0$
\State $i \gets 0$
\State LogLikelihood $= -\infty$
\While{LogLikelihood $< -\epsilon$}
\State E step: Set $A_\tau, B_\tau$ as per Equations \prettyref{eq:A} and \prettyref{eq:B}, assuming guesses $\sigma,\eta$
\State M step: Set $\sigma = \sqrt{2\sum_{\tau=1}^t A_\tau/t}, \eta = \sqrt{2 \sum_{\tau=1}^t B_\tau/t}$
\State LogLikelihood $\gets$ Equation \prettyref{eq:log_like1}
\State $i\gets i+1$
\EndWhile
\end{algorithmic}
\end{algorithm}

\subparagraph*{Managing computation, and adapting to a non-stationary market:} While Algorithm \prettyref{alg:akf} added on top of our AMM helps us estimate the unknown market parameters, its computational complexity keeps growing linearly with each additional trade. This is because every trade adds another term in the series that computes the log-likelihood function, hence increasing the number of iterations in both the forward and backward runs of the Kalman filter. Additionally, this algorithm also assumes there is no change in the market parameters $\eta,\sigma$ with time, that is, the market conditions are stationary. We can make the algorithm less computationally heavy and adapt to non-stationarities by truncating the trading data to a recent history of $p_{trad}^\tau$. This approach keeps track of all price estimates of the past, since they get refined with every backward run of the Adaptive Kalman algorithm, but only runs the algorithm on the recent history of price estimates and trading data.

\vspace{-0.1in}
\subsection{Adversarial robustness}
\label{sec:adverse}

A prominent danger to the proper functioning of market making protocols is the presence of adversarial trader behavior. The model described so far and the optimal solutions presented remain valid only when traders interacting with the market are rational with respect to the external market. However, there is always the possibility that an AMM is being manipulated for profits extracted from other protocols (e.g. lenders\cite{lendingAttacks}, derivative markets, etc.) relying on the AMM as a price oracle \cite{angeris2021constant}. Although any protocol using an AMM as a price oracle usually takes necessary precautions, such as ensuring a diverse portfolio of price signals, using outlier-robust statistics such as medians rather than means, etc. we show that our market making algorithms can be made robust to such market adversaries, when the proportion of such adversarial is less than half of all trading interactions.

\subparagraph*{Adversary model: }The adversarial behavior we seek to guard against is the manipulation of $p_{trad}^t$ that the AMM observes and not the external price $p_{ext}^t$. We assume that the external price is inferred from a deep market that is not easily manipulated. We further assume that a proportion $\alpha$ of the trader population is adversarial and the rest behave as per the rational model in \prettyref{sec:model}. However, since the AMM does not know which trades are being manipulated by the adversary, we assign a sequence of learnable weights $w_\tau$ for all trade observations in the past $\langle p_{trad}^\tau\rangle_{\tau=1}^t$. To successfully manipulate the price, the adversary needs to push $p_{trad}^t$ in a specific direction so as to induce a large discrepancy in the marginal prices of the AMM curve and $p_{ext}^t$. 

\subparagraph*{Robust adaptive curve algorithm: }A simple modification of the EM algorithm enables us to distinguish adversarial trades from honest trades \cite{chen2022kalman,ting2007}. We first rewrite the log-likelihood function of the AMM as
\begin{align}
     E[\log L_t | \langle p_{trad}^\tau\rangle_{\tau=1}^t] &= -\frac{t\log\sigma}{2} -\frac{t\log\eta}{2} + \sum_{\tau=1}^t \left(\frac{ \log w_\tau}{4}- \frac{ A_\tau}{2\sigma^2} - \frac{ w_\tau B_\tau}{2\eta^2}\right),\label{eq:log_like2}
\end{align}
which basically assumes that each datapoint has a different variance in noise $\eta^2/w_\tau$. This implies that data with low weights have a higher variance, and are hence the adversarial outliers. We start our algorithm with an equal weight given to all datapoints, and then estimate $A_\tau,B_\tau$ assuming those weights and running the forward and backward runs of the Kalman Filtering algorithm. After that, we set new weights by getting the critical points for the log-likelihood maximization using
\begin{align}
    \nabla_{w_\tau} E[\log L_t | \langle p_{trad}^\tau\rangle_{\tau=1}^t] = 0 \implies w_\tau^* &= \frac{\eta^2}{2B_\tau}.
\end{align}
This completes the adversarially robust version of the Kalman filtering algorithm. We empirically demonstrate the effectiveness of the approach for $\alpha < 0.5$ compared to static curves, and a naive Kalman filtering approach (\prettyref{sec:empirical}).

\section{Implications for AMMs with static curves}\label{sec:static_implications}

If we view the market maker as a price oracle, then one can compare the performance of different algorithms based on how the mean squared error of the AMM changes with the incoming trades. In particular, one can compare how quickly the error goes down in the adaptive protocol proposed in \prettyref{sec:akf} and a static curve such as Uniswap. This comparison has already been done for liquid markets \cite{nadkarni2023zeroswap}, where we observe an exponential decay of the error for an adaptive protocol compared to a linear decay for a static one. 

\subparagraph*{Error performance with trades: }For our case, let us assume that we are comparing how Algorithm \prettyref{alg:kf} performs in contrast to a static curve, if we assume that it is deployed on a blockchain with $T$ trades in a block. Then, we can prove that the error decreases with number of trades for Algorithm \prettyref{alg:kf} and stays constant for a static curve.

\begin{theorem}\label{thm:static_comparison}
    Let there be $T$ trades in a single block of transactions, with the external price at the creation of the block being $p_{ext}$. We denote by $p_{KF}^T$ and $p_{SC}^T$ the marginal prices of the Algorithm \prettyref{alg:kf} and a static curve at the end of the block. Then, we have
    \begin{align}
        E[(p_{ext}-p_{KF}^T)^2] &= \frac{\eta^2\sigma^2}{T\sigma^2 + \eta^2}\\
        E[(p_{ext}-p_{SC}^T)^2] &= \eta^2
    \end{align}
\end{theorem}


    

\subparagraph*{Implied dynamics of static curves: }We see that static curves are worse oracles because they do not use the realistic dynamical model to get the best estimate of the external price. The question then arises if there is any dynamical model that a particular  static curve \textit{is} optimal for. Note that the differential equation \prettyref{eq:main1} can be viewed as an equation in $\beta_t(p)$ if the curve $g_t(p)$ is given. We now use this observation to work out the \textit{implied} dynamical model underlying commonly used static curves. More formally, given the demand curve of a static AMM $g(p)$, we can find the corresponding $\beta_t(p)$ function by solving the following differential equation
\begin{align}
    \beta_t'(p) (g_t(p) - x_0^t) +\beta_t(p)g_t'(p)-pg_t'(p) &= 0 \label{eq:inverse1}
\end{align}
with the initial operating point $p_0^t$ satisfying the constraint $p=\beta_t(p)$. We now solve this equation for some common CFMM curves to get the underlying implied price/trader dynamics.

\subparagraph*{Constant Sum Market Makers} The demand curve for a constant sum market maker is given by 
\begin{align}
    g_t(p) = \begin{cases}
        x_0^t + y_0^t/p_0^t \ \ &\mathrm{if\ } p\leq p_0^t\\
        0 \ \ &\mathrm{if\ } p> p_0^t\\
    \end{cases}
\end{align}
where $p_0^t = p_0$ stays constant with time. Note that substituting the demand curve in \prettyref{eq:inverse1} gives a trivial equation $\beta_t'(p) = 0$. This, coupled with the condition on the initial operating point implies that $\beta_t(p) = p_0$ for all $t$. Clearly, this implies a dynamical model where $\Delta p_{ext}^t = \Delta p_{trad}^t = 0$. In other words, a CSMM assumes that the external price stays constant with time and that the traders have a noiseless view of the price at all times.

\subparagraph*{Constant Mean Market Makers} The demand curve for Constant Mean Market Makers is given by 
\begin{align}
    g_t(p) = \frac{c}{p^{1-\theta}}\left(\frac{\theta}{1-\theta}\right)^{1-\theta}
\end{align}
where the bonding curve for the constant mean market maker is of the form $x^\theta y^{1-\theta} = constant$, where $x,y$ denote the reserves for the asset and numeraire respectively.

Substituting this in \prettyref{eq:inverse1} and solving for $\beta_t(p)$ gives us the following function
\begin{align}
    \beta_t(p) &= \frac{1-\theta}{\theta} p^\theta(p_0^t)^{1-\theta} \frac{1-(p_0^t/p)^\theta}{1 - (p_0^t/p)^{1-\theta}}\label{eq:belief1}
\end{align}
which also satisfies $\beta_t(p_0^t) = p_0^t$. In particular, if we look at constant product market makers, we have $\theta=1/2$, giving us $\beta_t(p) = \sqrt{p_0^tp}$. For this $\beta_t(.)$ function, we have the following result.

\begin{theorem}\label{thm:static_implied}
    The following price and trader behavior dynamics yields $\beta_t(p)$ which obeys \prettyref{eq:belief1}. Equivalently, the following model has a constant mean market maker with weight parameter $\theta$ as its optimally efficient solution,
    \begin{align}
        \log p_{ext}^t &= \log p_{trad}^{t-1} + \epsilon_\sigma^t\label{eq:5update1}\\
        \log p_{trad}^t &= \log p_{ext}^{t} + \epsilon_\eta^t, \label{eq:5update2}
    \end{align}
    where $\epsilon_\sigma,\epsilon_\eta$ are independent Gaussian random variables with zero mean and variances $\sigma, \eta$ respectively, where the variances satisfy the following conditions.
    \begin{align}
    \sigma &\ll 1 \label{eq:5cond1}\\
        \eta &= \sigma\left(\sqrt{1/\theta-1}\right)\label{eq:5cond2}
    \end{align}
\end{theorem}

The above theorem sheds light on why static curves fail to prevent arbitrage loss as effectively - the implied dynamic model that they assume is mismatched with more realistic trader behavior. This mismatch manifests itself in the following ways, as indicated by \prettyref{eq:5cond1} and \prettyref{eq:5cond2}. 

\subparagraph*{Low latency blockchain: }Firstly, static curves assume that the price jumps between consecutive trades have a very small variance. This is equivalent to assuming that the inter-block times on the underlying blockchain go to $0$. This is because the standard deviation of price jumps $\sigma$ between blocks depends on the price volatility $\sigma'$ as $\sigma =\sigma'\sqrt{\Delta t}$, where $\Delta t$ is the inter-block arrival time. This observation confirms the conclusion reached in \cite{milionis2023automated} and the broader DeFi community \cite{lowLatencyLVR}, where the arbitrage loss (or LVR) is calculated for the special case of all of the trader population being arbitrageurs and concludes that this loss indeed approaches $0$ as the inter-block time goes to $0$.

\subparagraph*{Constraints on noise traders: }Secondly, the noise in the price that the traders see obeys a specific structure - $\eta^2 = \sigma^2 (1/\theta - 1)$. For a constant product market maker, this specifically assumes that the variance of price jumps is exactly the same as the variance of the noise in trader beliefs about the price. This implies additional restrictions on price and trader behaviour that might not always be true in real markets. However, these constraints can potentially help decide how toxic and non-toxic trade flow is guided in ``DEX-aggregator'' such as UniswapX \cite{uniswapX}, 1inch \cite{oneinch}, CoWSwap \cite{cowswap}, etc. to ensure that passive LPs (that invest in pools) get a fair price as defined by \prettyref{eq:gm_price_modified}. The main threat that such aggregators pose to passive LPs is that much of the non-toxic (or uninformed, or noise) trades get satisfied internally without any trades flowing through the pools, while the surplus (usually toxic or informed) gets routed though the passive pools \cite{dangerUniX}. The equation \prettyref{eq:5cond1} prescribes how much non-toxic flow should be ``added in the mix'' to ensure fairness to the passive pool LPs. 

\subparagraph*{Dominance of DEXes: }Thirdly, \prettyref{eq:5update1} highlights another key assumption - \textit{the external market also reacts to the price on the AMM}. This is only true when the AMM has an amount of liquidity that is more than the external market, which is not true for most AMM pools or decentralized exchanges today \cite{cexVSdex}. This means that static curves are guaranteed to make a loss to arbitrageurs unless decentralized exchanges become the main sources of liquidity, and the inter-block arrival time on the underlying blockchain become negligible.

\vspace{-0.1in}
\section{Empirical Results}\label{sec:empirical}

In this section, we present the empirical performance of the algorithms discussed so far in this work\footnote{Code for running all experiments has been shared \href{https://anonymous.4open.science/r/AdaptiveCurves-AFT24/}{\textit{here}}}.
\begin{figure}[hbt!]

\begin{subfigure}[t]{0.45\textwidth}
  \centering
\hspace*{-0.25in}
 \includegraphics[width=1.2\textwidth]{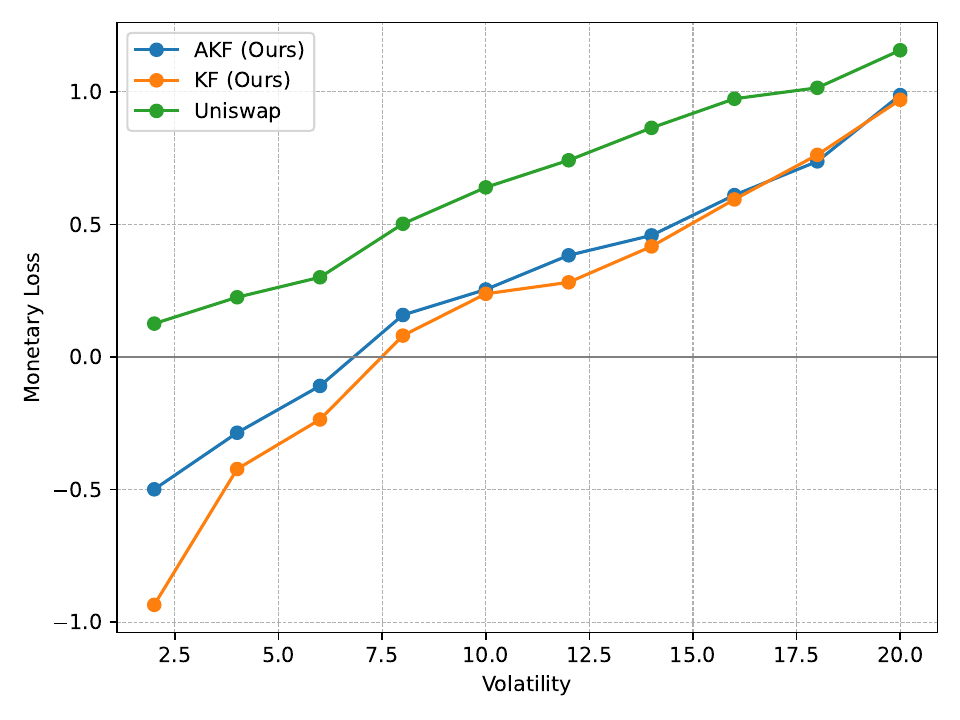}
  \caption{Varying price volatility $\sigma$}
  \label{fig:variab_loss_avg_vs_sigma}
\end{subfigure}
\hfill
\begin{subfigure}[t]{0.45\textwidth}
  \centering
\hspace*{-0.25in}
 \includegraphics[width=1.2\textwidth]{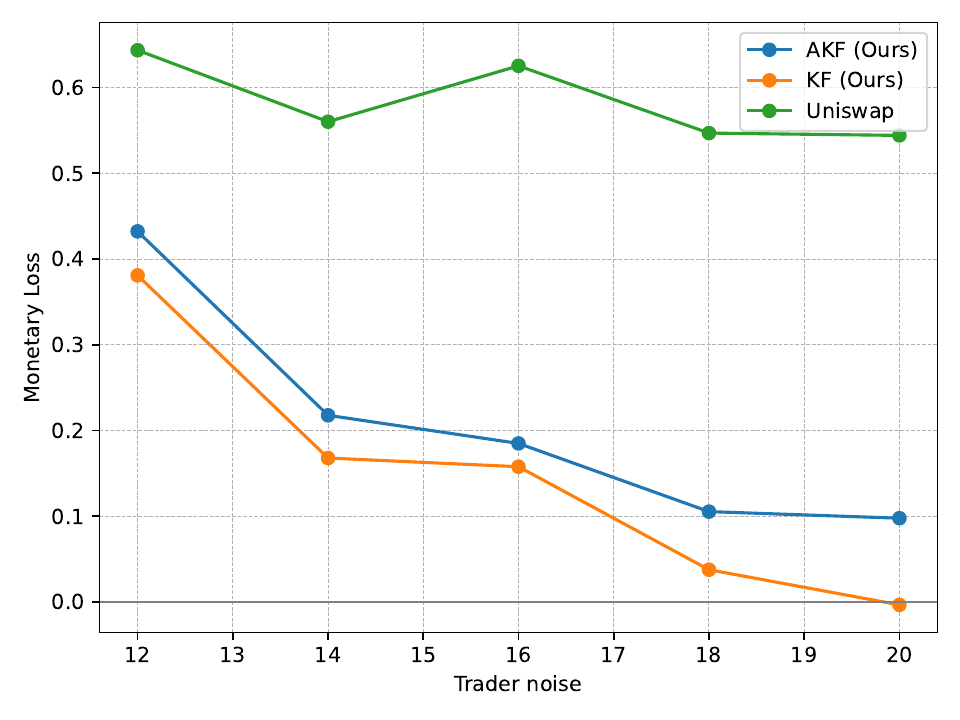}
  \caption{Varying trader noise $\eta$}
  \label{fig:variab_spread_avg_vs_sigma}
\end{subfigure}

\caption{Percentage monetary loss per trade of our market making algorithms (Kalman Filtering and Adaptive Kalman Filtering) is much less than a static Uniswap curve for a Gaussian price jump and trader noise models}
\label{fig:variab_monetary_loss}
\end{figure}

 \subparagraph*{Comparison with static curves} We simulate the model described in \prettyref{sec:model} and compare the performance of the algorithms we proposed (\prettyref{fig:variab_monetary_loss}). We see that adapting the AMM curve according to algorithms \prettyref{alg:kf} and \prettyref{alg:akf} give a much lower monetary loss per trade than a static constant product curve that is used in Uniswap. Furthermore, the Adaptive  Kalman Filter algorithm estimates the unknown market parameters correctly, leading to it achieving close performance with the optimal Kalman Filtering algorithm.

\begin{figure}[hbt!]

\begin{subfigure}[t]{0.45\textwidth}
  \centering
\hspace*{-0.25in}
 \includegraphics[width=1.2\textwidth]{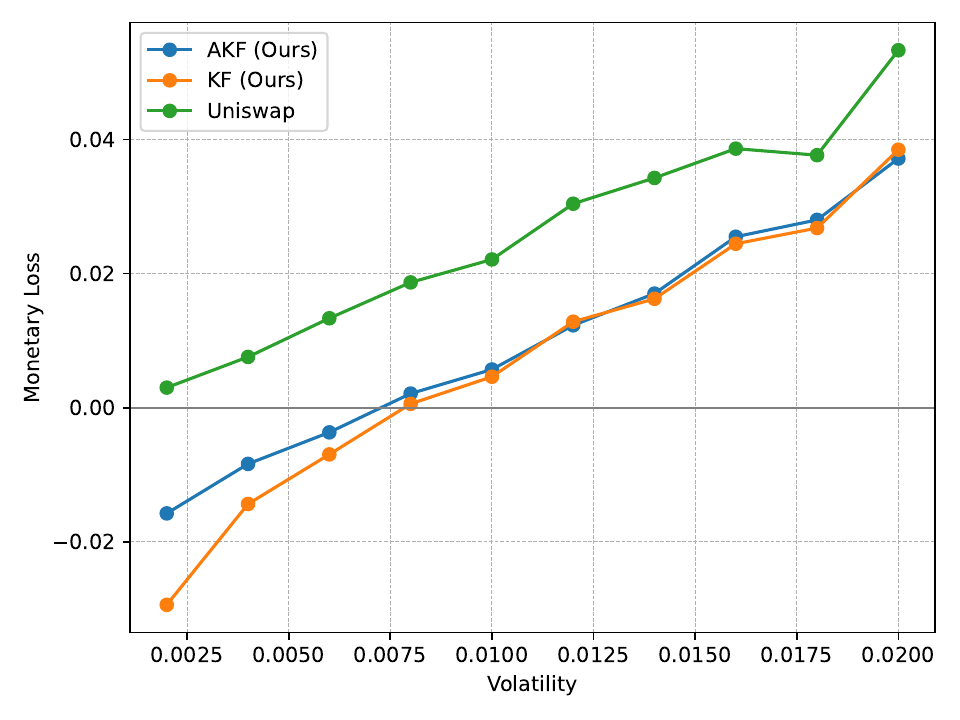}
  \caption{Varying price volatility $\sigma$}
  \label{fig:variab_loss_avg_vs_sigma_gbm}
\end{subfigure}
\hfill
\begin{subfigure}[t]{0.45\textwidth}
  \centering
\hspace*{-0.25in}
 \includegraphics[width=1.2\textwidth]{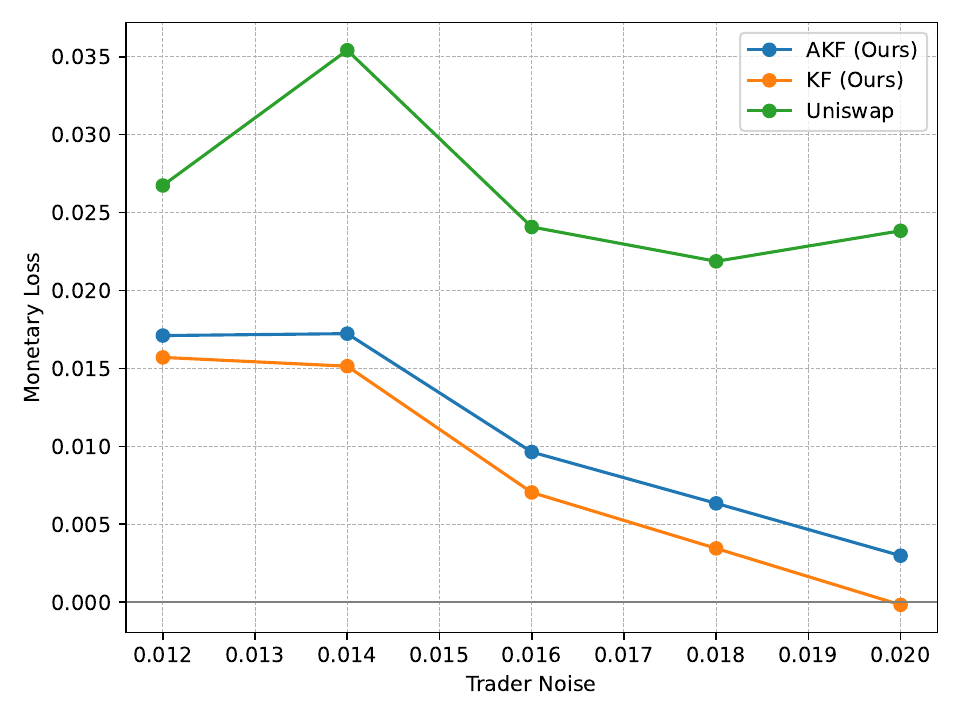}
  \caption{Varying trader noise $\eta$}
  \label{fig:variab_spread_avg_vs_sigma_gbm}
\end{subfigure}

\caption{Percentage monetary loss per trade of our algorithms (Kalman Filtering and Adaptive Kalman Filtering) is much less than a static Uniswap curve for a Lognormal price jump and trader noise models}
\label{fig:variab_monetary_loss_gbm}
\end{figure}

The same observation holds for prices that follow a geometric Brownian motion, as seen in \prettyref{fig:variab_monetary_loss_gbm}.

\begin{figure}[hbt!]

\begin{subfigure}[t]{0.45\textwidth}
  \centering
\hspace*{-0.25in}
 \includegraphics[width=1.2\textwidth]{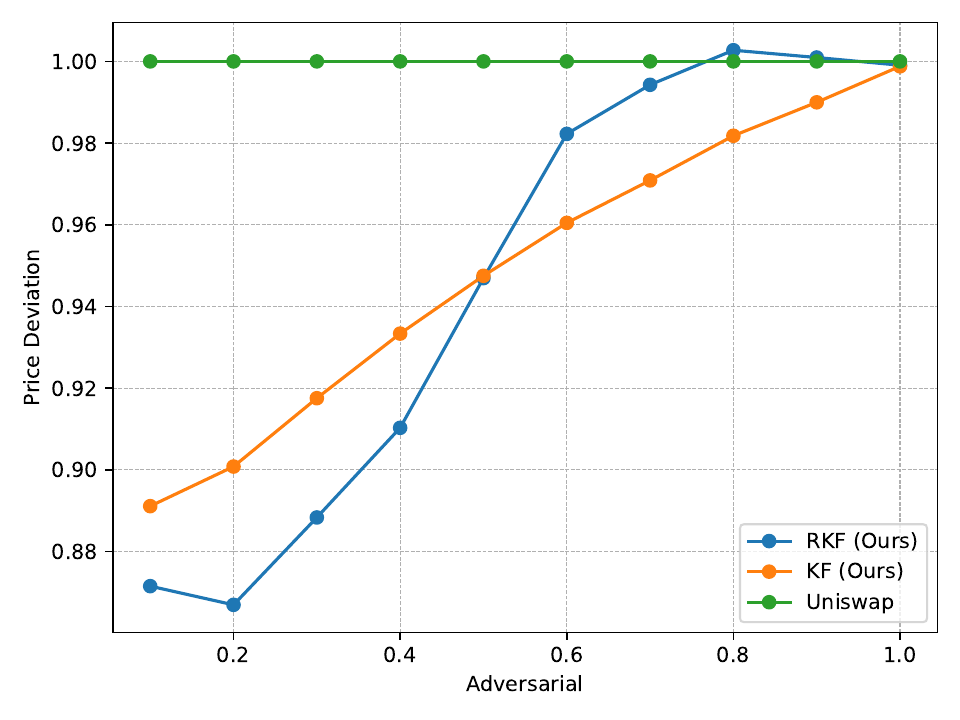}
  \caption{Root Mean Squared Deviation of external price from the price recommended by the AMM}
  \label{fig:variab_loss_avg_vs_sigma_adv}
\end{subfigure}
\hfill
\begin{subfigure}[t]{0.45\textwidth}
  \centering
\hspace*{-0.25in}
 \includegraphics[width=1.2\textwidth]{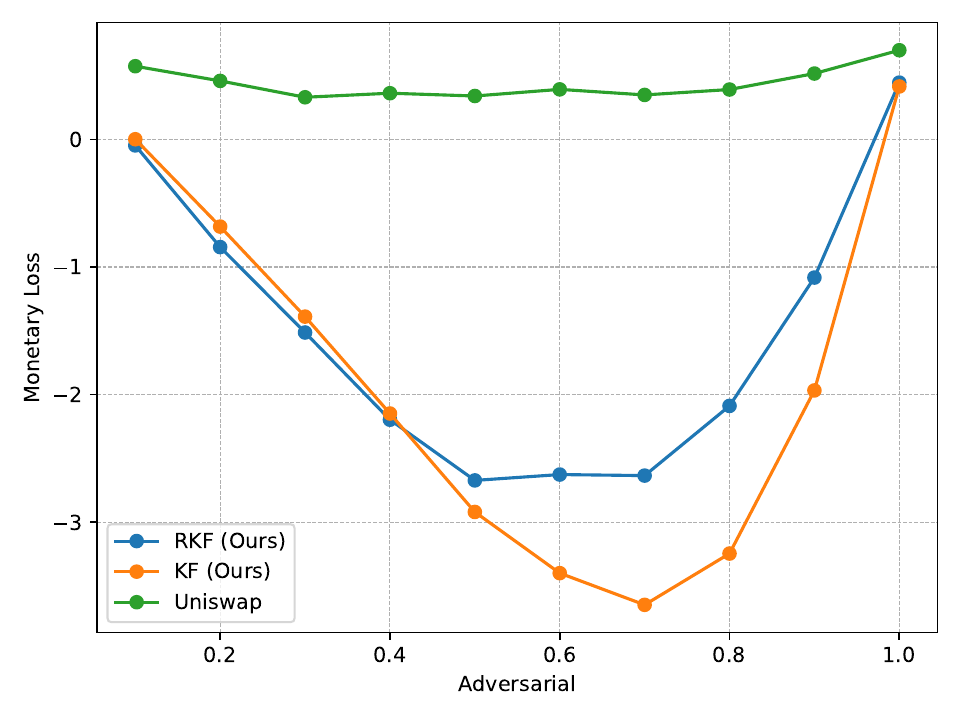}
  \caption{Percentage monetary loss compared across AMMs}
  \label{fig:variab_spread_avg_vs_sigma_adv}
\end{subfigure}

\caption{If the AMM is treated as an oracle, then we can use the Robust Kalman Filtering to get a more accurate reading of the hidden external price}
\label{fig:variab_monetary_loss_adv}
\end{figure}

\subparagraph*{Robustness to adversarial traders} In presence of adversarial traders, the Robust Kalman Filtering algorithm is used to change the curve of the AMM gives us a more accurate reading of the hidden external price in the presence of less than $50\%$ of the population of traders being adversarial (\prettyref{fig:variab_monetary_loss_adv}). The adversarial traders, in this case, are assumed to perform large buy trades (with price belief $p_{trad}^t$ about 5-7 standard deviations beyond normal trading size) to keep the AMM price above the external market. We also note that the monetary loss of the AMMs against the adversary stays in the profitable region for the adaptive curves, while the static curve suffers a loss to arbitrageurs even if the adversary is making trades that are irrational.

\begin{figure}[hbt!]

\begin{subfigure}[t]{0.45\textwidth}
  \centering
\hspace*{-0.25in}
 \includegraphics[width=1.2\textwidth]{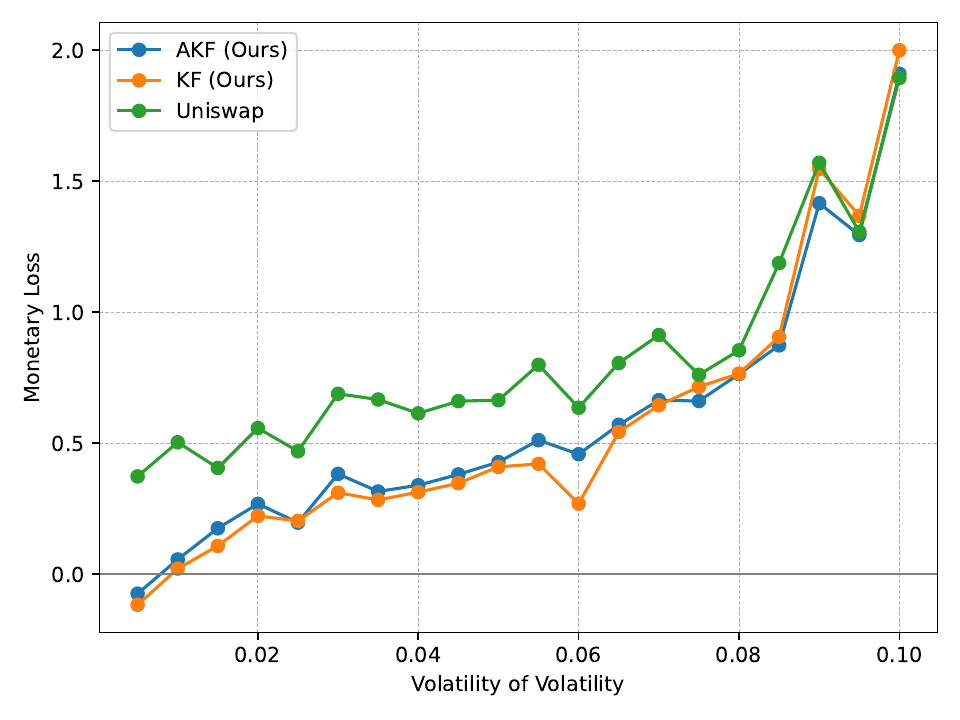}
  \caption{Varying Volatility of Volatility}
  \label{fig:variab_params1}
\end{subfigure}
\hfill
\begin{subfigure}[t]{0.45\textwidth}
  \centering
\hspace*{-0.25in}
 \includegraphics[width=1.2\textwidth]{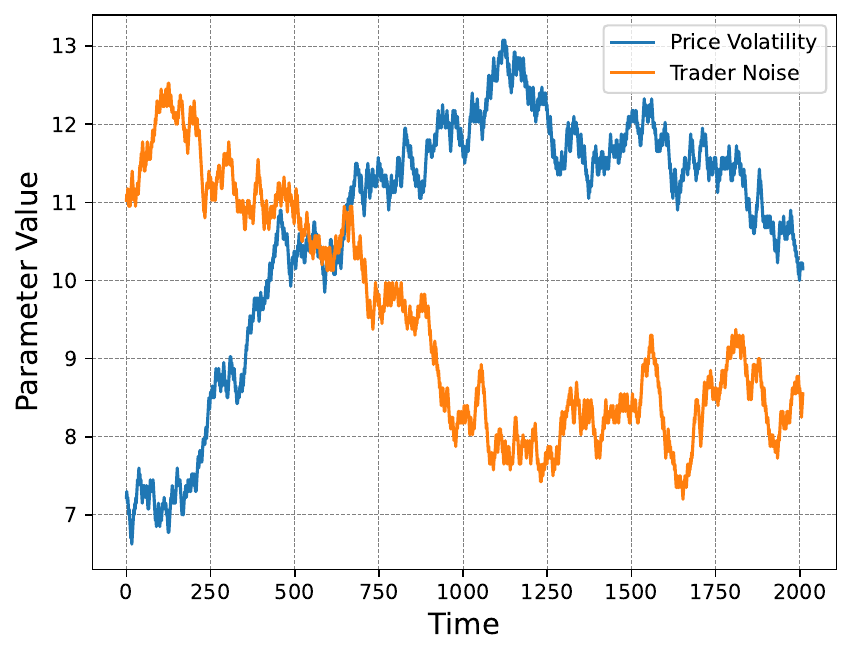}
  \caption{Changing market conditions $\eta,\sigma$ for a single sample path}
  \label{fig:variab_params2}
\end{subfigure}

\caption{Percentage monetary loss per trade of our market making algorithms (Kalman Filtering and Adaptive Kalman Filtering) is much less than a static Uniswap curve for continously changing market conditions}
\label{fig:variab_params}
\end{figure}

 \subparagraph*{Robustness to non-stationary markets} \prettyref{fig:variab_params} tests the truncated version of the adaptive Kalman algorithm for changing market conditions. This algorithm is compared with the Kalman Filter algorithm, that knows the underlying market parameters exactly at every time step, and the static Uniswap curve. The monetary loss incurred by the AMMs is measured against changing variability of market conditions, measured by the volatility of price jump variance and trader noise (termed as volatility of volatility $\sigma_{\eta,\sigma}$). This metric measures the standard deviation of changing $\eta,\sigma$ after every trade, with a sample path of these changing variables shown in \prettyref{fig:variab_params2} for $\sigma_{\eta,\sigma} = 0.04$. Recall that $\eta, \sigma$ themselves govern how the external price is percieved by the traders and how it changes after every trade respectively. We see that the Adaptive Kalman Filter is able to perform almost as well as the Kalman Filter when the market changes are slow enough. However, the performance advantage over static curves vanishes as the changing market conditions become more erratic. This happens because the timescale over which market parameters suffer large changes becomes comparable to the timescale of the recent history considered by the truncated adaptive Kalman filter. This observation offers guidance to AMM designers on choosing the timescale over which adaptivity can offer an advantage over static markets.

\vspace{-0.1in}
\section{On-chain System Implementation}
\label{sec:implementation}
\begin{figure}[hbt!]
  \centering
\hspace*{-0.4in}
 \includegraphics[width=0.98\textwidth]{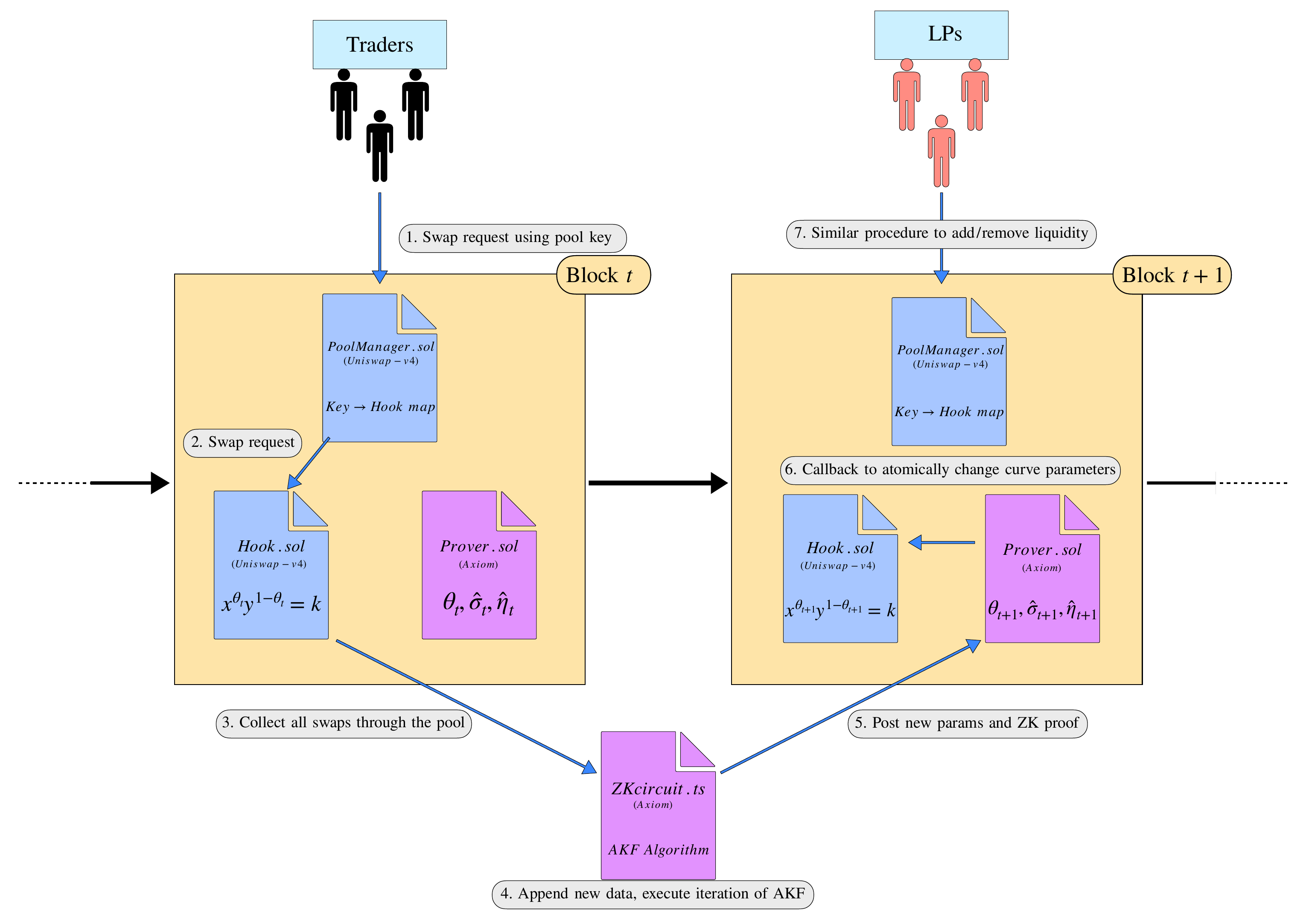}
\caption{System design for an on-chain implementation of our algorithms}
\label{fig:implem}
\end{figure}

Many prior works \cite{churiwala2022qlammp, nadkarni2023zeroswap} seek to implement adaptive market makers on a blockchain, where the adapting is done using machine learning algorithms that must be necessarily performed off-chain because of their computational load \cite{offchainCompute}. To that end, a group of protocol validators (separate from the validators of the underlying blockchain) are assumed, who run the bulk of the computation off-chain and post their results (such as satisfied orders, their prices, etc.) on-chain. However, recent developments in Layer 2 or rollup \cite{arbitrum, optimism} infrastructure, machine learning co-processors with zero-knowledge guarantees \cite{axiom,brevis}, has given rise to several platforms that can be utilized directly to implement the adaptive market makers (or machine learning algorithms in general) we derived in the previous sections, without any need of additional validators. We draw upon these innovations for the blockchain implementation, and divide the approach into two parts. The overall design has been shown in \prettyref{fig:implem}.

\subparagraph*{Hook Contract: }The first part of such an implementation is the liquidity pool contract, which allows the canonical interactions with LPs and traders given a specific demand/bonding curve. In the Ethereum DeFi ecosystem, a recent proposal \cite{uniswapv4} by the Uniswap protocol (called Uniswap-v4) presents a highly customizable platform for adaptive market making. The main innovation is the introduction of a ``hook'' smart contract \cite{hooksUniv4}. While prior versions \cite{uniswapv2,uniswapv3} of the protocol presented market making in a canonical manner where the exact bonding curve followed during trade execution was static and only gave freedoms to LPs in terms of the distribution of liquidity along that fixed curve, the new platform allows the LP to specify changes to the bonding curve just before/after every single trade via the hook contract. We use this functionality to execute trades and add/remove liquidity according to the curve $x^{\theta_t}y^{1-\theta_t} = constant$. When traders/LPs put in canonical swap/liquidity addition transactions, they are first routed through the \texttt{PoolManager} contract. This maps a pool to a specific \texttt{Hook} contract that specifies all modifications to the curve before/after trade execution. All interactions with this contract are collected as input data by the second part of the system.

\subparagraph*{Off-Chain Co-processor: }The second part runs the algorithm used to change the demand/bonding curve given the history of trader interactions. This is done off-chain due to the high computational load of running the EM algorithm as part of Adaptive Kalman Filtering. For our implementation, we chose Axiom \cite{axiom} as a platform to run this off-chain computation. The main part of this implementation is a typescript file \texttt{ZKcircuit} containing the details of Algorithm \prettyref{alg:akf} implemented as a algebraic circuit, so that a zero-knowledge proof can be generated corresponding to the computation \cite{axiomCircuit}. This file also verifiably collects data from the previous block, and runs the algorithm to come up with new estimates for the market parameters $\sigma, \eta$  and hence the curve parameter $\theta$. The off-chain Axiom client \cite{axiomCircuit} posts this in the next block and is verified by the on-chain \texttt{Prover} contract of Axiom. This also invokes a callback to the \texttt{Hook} where the changes to the curve parameters are finally implemented. We provide open access to all files used in our implementation \footnote{Code for the proof-of-concept implementation has been shared \href{https://anonymous.4open.science/r/AdaptiveCurves-AFT24/}{\textit{here}}}.


\section{Conclusion and Future Work}
\label{sec:conclusion}
\color{black}

\subparagraph*{Generalising to non-Gaussian and non-stationary behavior: }Traders usually do not perfectly conform to the distributional assumptions we use to derive optimal solutions in \prettyref{sec:kf}. Our approach can be potentially be extended to such situations by the use of Neural Kalman Filters \cite{millidge2021neural}, which claim to work for non-Gaussian/non-stationary state space models. 

\subparagraph*{Balancing toxic and non-toxic orderflow: }The blockchain-level conditions for the optimality of static curves, as outlined in \prettyref{sec:static_implications}, provide guidance on how toxic/non-toxic orderflow, if discriminated correctly \cite{lowLatencyLVR}, should be allowed to use passive liquidity and still give LPs a fair price. Developing DEX aggregators that aware of these conditions would help limit the dangers to passive LPs in DeFi.

\subparagraph*{Extensions to other adaptive protocols: }In this work, we have derived a correspondence between a dynamical model for prices and its optimal market making curve. This principle can be extended for stable control of other DeFi protocols, such as lending, that currently use static curves.

\bibliography{references}

\begin{thebibliography}{10}

\bibitem{oneinch}
1inch protocol.
\newblock 1inch dex aggreagator.
\newblock \url{https://1inch.io}.
\newblock Accessed: 2023-05.

\bibitem{angeris2021constant}
Guillermo Angeris, Akshay Agrawal, Alex Evans, Tarun Chitra, and Stephen Boyd.
\newblock Constant function market makers: Multi-asset trades via convex optimization, 2021.
\newblock \href {https://arxiv.org/abs/2107.12484} {\path{arXiv:2107.12484}}.

\bibitem{Angeris_2020}
Guillermo Angeris and Tarun Chitra.
\newblock Improved price oracles.
\newblock In {\em Proceedings of the 2nd {ACM} Conference on Advances in Financial Technologies}. {ACM}, oct 2020.
\newblock URL: \url{https://doi.org/10.1145%2F3419614.3423251}, \href {https://doi.org/10.1145/3419614.3423251} {\path{doi:10.1145/3419614.3423251}}.

\bibitem{angeris2020does}
Guillermo Angeris, Alex Evans, and Tarun Chitra.
\newblock When does the tail wag the dog? curvature and market making, 2020.
\newblock \href {https://arxiv.org/abs/2012.08040} {\path{arXiv:2012.08040}}.

\bibitem{aoyagi2020LP}
Jun Aoyagi.
\newblock Liquidity provision by automated market makers, 2020.
\newblock URL: \url{https://ssrn.com/abstract=3674178}.

\bibitem{lendingAttacks}
Ayana~T. Aspembitova and Michael~A. Bentley.
\newblock Oracles in decentralized finance: Attack costs, profits and mitigation measures.
\newblock {\em Entropy}, 25(1), 2023.
\newblock URL: \url{https://www.mdpi.com/1099-4300/25/1/60}, \href {https://doi.org/10.3390/e25010060} {\path{doi:10.3390/e25010060}}.

\bibitem{avellaneda2008}
Marco Avellaneda and Sasha Stoikov.
\newblock High frequency trading in a limit order book.
\newblock {\em Quantitative Finance}, 8:217--224, 04 2008.
\newblock \href {https://doi.org/10.1080/14697680701381228} {\path{doi:10.1080/14697680701381228}}.

\bibitem{dangerUniX}
Ashwath Balakrishnan.
\newblock Understanding uniswap's new liquidity aggregator (tl;dr at the end).
\newblock \url={https://members.delphidigital.io/feed/understanding-uniswaps-new-liquidity-aggregator-tldr-at-the-end}, 2023.
\newblock \href {https://arxiv.org/abs/2305.14604} {\path{arXiv:2305.14604}}.

\bibitem{tradeVolComparisons}
The Block.
\newblock Block.co trade volume comparisons.
\newblock \url{https://www.theblock.co/data/decentralized-finance/dex-non-custodial/uniswap-vs-coinbase-and-binance-trade-volume-7dma}.
\newblock Accessed: 2023-09.

\bibitem{brevis}
Brevis.
\newblock Brevis website.
\newblock \url{https://docs.brevis.network/ }.
\newblock Accessed: 2023-05.

\bibitem{cartea24strategic}
Álvaro Cartea, Fayçal Drissi, Leandro Sánchez-Betancourt, David Siska, and Lukasz Szpruch.
\newblock Strategic bonding curves in automated market makers.
\newblock 2024.
\newblock URL: \url{https://ssrn.com/abstract=5018420}.

\bibitem{cexVSdex}
centicio.
\newblock Centralized exchange (cex) vs decentralized exchange (dex). which is the best crypto exchange?
\newblock \url{https://medium.com/@centicio/centralized-exchange-cex-vs-decentralized-exchange-dex-which-is-the-best-crypto-exchange-148f48ea51c1}.
\newblock Accessed: 2023-05.

\bibitem{chan2001}
Nicholas Chan and Christian Shelton.
\newblock An electronic market-maker, 01 2001.

\bibitem{chen2022kalman}
Sitan Chen, Frederic Koehler, Ankur Moitra, and Morris Yau.
\newblock Kalman filtering with adversarial corruptions.
\newblock In {\em Proceedings of the 54th Annual ACM SIGACT Symposium on Theory of Computing}, pages 832--845, 2022.

\bibitem{churiwala2022qlammp}
Dev Churiwala and Bhaskar Krishnamachari.
\newblock Qlammp: A q-learning agent for optimizing fees on automated market making protocols, 2022.
\newblock \href {https://arxiv.org/abs/2211.14977} {\path{arXiv:2211.14977}}.

\bibitem{crocToxicity}
CrocSwap.
\newblock Discrimination of toxic flow in uniswap v3.
\newblock \url{https://crocswap.medium.com/discrimination-of-toxic-flow-in-uniswap-v3-part-1-fb5b6e01398b}.
\newblock Accessed: 2023-09.

\bibitem{dodoOracle}
Diane Dai.
\newblock Dodo integrates chainlink live on mainnet, kickstarts the on-chain liquidity revolution.
\newblock \url{https://blog.dodoex.io/dodo-integrates-chainlink-live-on-mainnet-kickstarts-the-on-chain-liquidity-revolution-ee27e136e122}.
\newblock Accessed: 2023-09.

\bibitem{das2005learning}
Sanmay Das*.
\newblock A learning market-maker in the glosten--milgrom model.
\newblock {\em Quantitative Finance}, 5(2):169--180, 2005.

\bibitem{das2008adapting}
Sanmay Das and Malik Magdon-Ismail.
\newblock Adapting to a market shock: Optimal sequential market-making.
\newblock {\em Advances in Neural Information Processing Systems}, 21, 2008.

\bibitem{emog79}
A.~P. Dempster, N.~M. Laird, and D.~B. Rubin.
\newblock Maximum likelihood from incomplete data via the em algorithm.
\newblock {\em Journal of the Royal Statistical Society. Series B (Methodological)}, 39(1):1--38, 1977.
\newblock URL: \url{http://www.jstor.org/stable/2984875}.

\bibitem{elliott2006forecasting}
Graham Elliott.
\newblock Forecasting with trending data.
\newblock {\em Handbook of economic forecasting}, 1:555--604, 2006.

\bibitem{Eskandari_2021}
Shayan Eskandari, Mehdi Salehi, Wanyun~Catherine Gu, and Jeremy Clark.
\newblock {SoK}.
\newblock In {\em Proceedings of the 3rd {ACM} Conference on Advances in Financial Technologies}. {ACM}, sep 2021.
\newblock URL: \url{https://doi.org/10.1145%2F3479722.3480994}, \href {https://doi.org/10.1145/3479722.3480994} {\path{doi:10.1145/3479722.3480994}}.

\bibitem{evans2020liquidity}
Alex Evans.
\newblock Liquidity provider returns in geometric mean markets, 2020.
\newblock \href {https://arxiv.org/abs/2006.08806} {\path{arXiv:2006.08806}}.

\bibitem{evans2021optimal}
Alex Evans, Guillermo Angeris, and Tarun Chitra.
\newblock Optimal fees for geometric mean market makers, 2021.
\newblock \href {https://arxiv.org/abs/2104.00446} {\path{arXiv:2104.00446}}.

\bibitem{frongillo2023axiomatic}
Rafael Frongillo, Maneesha Papireddygari, and Bo~Waggoner.
\newblock An axiomatic characterization of cfmms and equivalence to prediction markets.
\newblock {\em arXiv preprint arXiv:2302.00196}, 2023.

\bibitem{GLOSTEN198571}
Lawrence~R. Glosten and Paul~R. Milgrom.
\newblock Bid, ask and transaction prices in a specialist market with heterogeneously informed traders.
\newblock {\em Journal of Financial Economics}, 14(1):71--100, 1985.
\newblock URL: \url{https://www.sciencedirect.com/science/article/pii/0304405X85900443}, \href {https://doi.org/10.1016/0304-405X(85)90044-3} {\path{doi:10.1016/0304-405X(85)90044-3}}.

\bibitem{goyal2023finding}
Mohak Goyal, Geoffrey Ramseyer, Ashish Goel, and David Mazières.
\newblock Finding the right curve: Optimal design of constant function market makers, 2023.
\newblock \href {https://arxiv.org/abs/2212.03340} {\path{arXiv:2212.03340}}.

\bibitem{grossmanMiller}
SANFORD~J. GROSSMAN and MERTON~H. MILLER.
\newblock Liquidity and market structure.
\newblock {\em The Journal of Finance}, 43(3):617--633, 1988.
\newblock URL: \url{https://onlinelibrary.wiley.com/doi/abs/10.1111/j.1540-6261.1988.tb04594.x}, \href {https://arxiv.org/abs/https://onlinelibrary.wiley.com/doi/pdf/10.1111/j.1540-6261.1988.tb04594.x} {\path{arXiv:https://onlinelibrary.wiley.com/doi/pdf/10.1111/j.1540-6261.1988.tb04594.x}}, \href {https://doi.org/10.1111/j.1540-6261.1988.tb04594.x} {\path{doi:10.1111/j.1540-6261.1988.tb04594.x}}.

\bibitem{gustafsson2002particle}
F.~Gustafsson, F.~Gunnarsson, N.~Bergman, U.~Forssell, J.~Jansson, R.~Karlsson, and P.-J. Nordlund.
\newblock Particle filters for positioning, navigation, and tracking.
\newblock {\em IEEE Transactions on Signal Processing}, 50(2):425--437, 2002.
\newblock \href {https://doi.org/10.1109/78.978396} {\path{doi:10.1109/78.978396}}.

\bibitem{Heimbach_2022}
Lioba Heimbach, Eric Schertenleib, and Roger Wattenhofer.
\newblock Risks and returns of uniswap v3 liquidity providers.
\newblock In {\em Proceedings of the 4th {ACM} Conference on Advances in Financial Technologies}. {ACM}, sep 2022.
\newblock URL: \url{https://doi.org/10.1145%2F3558535.3559772}, \href {https://doi.org/10.1145/3558535.3559772} {\path{doi:10.1145/3558535.3559772}}.

\bibitem{hoAndStoll}
Thomas S.~Y. Ho and Hans~R. Stoll.
\newblock The dynamics of dealer markets under competition.
\newblock {\em The Journal of Finance}, 38(4):1053--1074, 1983.
\newblock URL: \url{http://www.jstor.org/stable/2328011}.

\bibitem{stablecoin3}
Peter Johnson and Sai Nimmagadda.
\newblock The relentless rise of stablecoins.
\newblock \url{https://digify.com/a/#/f/p/ef09be008ee64ab68bda4f0a558302a2}.
\newblock Accessed: 2023-09.

\bibitem{kalman}
Rudolph~Emil Kalman.
\newblock A new approach to linear filtering and prediction problems.
\newblock {\em Transactions of the ASME--Journal of Basic Engineering}, 82(Series D):35--45, 1960.

\bibitem{arbitrum}
Harry Kalodner, Steven Goldfeder, Xiaoqi Chen, S.~Matthew Weinberg, and Edward~W. Felten.
\newblock Arbitrum: Scalable, private smart contracts.
\newblock In {\em 27th USENIX Security Symposium (USENIX Security 18)}, pages 1353--1370, Baltimore, MD, August 2018. USENIX Association.
\newblock URL: \url{https://www.usenix.org/conference/usenixsecurity18/presentation/kalodner}.

\bibitem{kyleModel}
Albert~S. Kyle.
\newblock Continuous auctions and insider trading.
\newblock {\em Econometrica}, 53(6):1315--1335, 1985.
\newblock URL: \url{http://www.jstor.org/stable/1913210}.

\bibitem{univ3}
DeFi Llama.
\newblock Uniswap-v3 tvl comparison for stable coins vs non-stablecoins.
\newblock \url{https://defillama.com/protocol/uniswap-v3}.
\newblock Accessed: 2023-09.

\bibitem{loesch2021impermanent}
Stefan Loesch, Nate Hindman, Mark~B Richardson, and Nicholas Welch.
\newblock Impermanent loss in uniswap v3, 2021.
\newblock \href {https://arxiv.org/abs/2111.09192} {\path{arXiv:2111.09192}}.

\bibitem{frontrunningOracles}
Christos Makidris.
\newblock Front running, bots, slippage, oracle pricing errors: Amms are great, but there are problems.
\newblock \url{https://cointelegraph.com/magazine/trouble-with-crypto-automated-market-makers/}.
\newblock Accessed: 2023-09.

\bibitem{balancerWP}
F.~Martinelli.
\newblock Balancer whitepaper.
\newblock \url{https://balancer.fi/whitepaper.pdf}.
\newblock Accessed: 2023-05.

\bibitem{mcmenamin2022diamonds}
Conor McMenamin, Vanesa Daza, and Bruno Mazorra.
\newblock Diamonds are forever, loss-versus-rebalancing is not, 2022.
\newblock \href {https://arxiv.org/abs/2210.10601} {\path{arXiv:2210.10601}}.

\bibitem{mehra1970adaptive}
Raman~K. Mehra.
\newblock On the identification of variances and adaptive kalman filtering.
\newblock {\em IEEE Transactions on Automatic Control}, 15:175--184, 1970.
\newblock URL: \url{https://api.semanticscholar.org/CorpusID:238574860}.

\bibitem{milionis2023automated}
Jason Milionis, Ciamac~C. Moallemi, and Tim Roughgarden.
\newblock Automated market making and arbitrage profits in the presence of fees, 2023.
\newblock \href {https://arxiv.org/abs/2305.14604} {\path{arXiv:2305.14604}}.

\bibitem{milionis2023myersonian}
Jason Milionis, Ciamac~C. Moallemi, and Tim Roughgarden.
\newblock A myersonian framework for optimal liquidity provision in automated market makers, 2023.
\newblock \href {https://arxiv.org/abs/2303.00208} {\path{arXiv:2303.00208}}.

\bibitem{milionis2022automated}
Jason Milionis, Ciamac~C. Moallemi, Tim Roughgarden, and Anthony~Lee Zhang.
\newblock Automated market making and loss-versus-rebalancing, 2022.
\newblock \href {https://arxiv.org/abs/2208.06046} {\path{arXiv:2208.06046}}.

\bibitem{millidge2021neural}
Beren Millidge, Alexander Tschantz, Anil Seth, and Christopher Buckley.
\newblock Neural kalman filtering, 2021.
\newblock \href {https://arxiv.org/abs/2102.10021} {\path{arXiv:2102.10021}}.

\bibitem{mohanDexPrimer}
Vijay Mohan.
\newblock Automated market makers and decentralized exchanges: a defi primer, 12 2020.
\newblock \href {https://doi.org/10.2139/ssrn.3722714} {\path{doi:10.2139/ssrn.3722714}}.

\bibitem{nadkarni2023zeroswap}
Viraj Nadkarni, Jiachen Hu, Ranvir Rana, Chi Jin, Sanjeev Kulkarni, and Pramod Viswanath.
\newblock Zeroswap: Data-driven optimal market making in defi.
\newblock {\em arXiv preprint arXiv:2310.09413}, 2023.

\bibitem{nadkarni2024adaptive}
Viraj Nadkarni, Sanjeev Kulkarni, and Pramod Viswanath.
\newblock Adaptive curves for optimally efficient market making.
\newblock {\em arXiv preprint arXiv:2406.13794}, 2024.

\bibitem{nezlobinToxicity}
Alex Nezlobin.
\newblock Order flow toxicity on dexes.
\newblock \url{https://ethresear.ch/t/order-flow-toxicity-on-dexes/13177}.
\newblock Accessed: 2023-09.

\bibitem{optimism}
Optimism.
\newblock Optimism docs.
\newblock \url{https://community.optimism.io/}.
\newblock Accessed: 2023-09.

\bibitem{axiomCircuit}
Axiom Protocol.
\newblock {Axiom} client ciruit.
\newblock \url{https://docs.axiom.xyz/docs/axiom-developer-flow/axiom-client-circuit}.
\newblock Accessed: 2023-05.

\bibitem{axiom}
Axiom Protocol.
\newblock {Axiom} website.
\newblock \url{https://docs.axiom.xyz/}.
\newblock Accessed: 2023-05.

\bibitem{cowswap}
CoW protocol.
\newblock Cowswap docs.
\newblock \url{https://docs.cow.fi/overview/coincidence-of-wants}.
\newblock Accessed: 2023-09.

\bibitem{offchainCompute}
Gate protocol].
\newblock Offchain compute is all you need.
\newblock \url{https://www.gate.io/learn/articles/off-chain-compute-is-alll-you-need/1225}.
\newblock Accessed: 2023-05.

\bibitem{hooksUniv4}
Uniswap Protocol.
\newblock Hooks on uniswap v4.
\newblock \url{https://docs.uniswap.org/contracts/v4/concepts/hook-deployment}.
\newblock Accessed: 2023-05.

\bibitem{uniswapv2}
Uniswap Protocol.
\newblock Uniswap v2 core.
\newblock \url{https://uniswap.org/whitepaper.pdf}.
\newblock Accessed: 2023-09.

\bibitem{uniswapv3}
Uniswap Protocol.
\newblock Uniswap v3 core.
\newblock \url{https://uniswap.org/whitepaper-v3.pdf}.
\newblock Accessed: 2023-09.

\bibitem{uniswapv4}
Uniswap Protocol.
\newblock Uniswap v4 docs.
\newblock \url{https://docs.uniswap.org/contracts/v4/concepts/intro-to-v4}.
\newblock Accessed: 2023-09.

\bibitem{uniswapX}
Uniswap Protocol.
\newblock Uniswapx docs.
\newblock \url{https://blog.uniswap.org/uniswapx-protocol}.
\newblock Accessed: 2023-09.

\bibitem{flashLoansAttack}
Palamarchuk Roman.
\newblock Flash loan attacks: Risks and prevention.
\newblock \url{https://hacken.io/discover/flash-loan-attacks/}.
\newblock Accessed: 2023-05.

\bibitem{shumway1982approach}
Robert~H Shumway and David~S Stoffer.
\newblock An approach to time series smoothing and forecasting using the em algorithm.
\newblock {\em Journal of time series analysis}, 3(4):253--264, 1982.

\bibitem{tangri2023generalizing}
Rohan Tangri, Peter Yatsyshin, Elisabeth~A. Duijnstee, and Danilo Mandic.
\newblock Generalizing impermanent loss on decentralized exchanges with constant function market makers, 2023.
\newblock \href {https://arxiv.org/abs/2301.06831} {\path{arXiv:2301.06831}}.

\bibitem{lowLatencyLVR}
TheiaResearch.
\newblock Better blockchains lead to more profitable liquidity providers.
\newblock \url{https://twitter.com/TheiaResearch/status/1790717593440952757}.
\newblock Accessed: 2024.

\bibitem{thrun2005probabilistic}
Sebastian Thrun, Wolfram Burgard, and Dieter Fox.
\newblock {\em Probabilistic robotics}.
\newblock MIT Press, Cambridge, Mass., 2005.
\newblock URL: \url{http://www.amazon.de/gp/product/0262201623/102-8479661-9831324?v=glance&n=283155&n=507846&s=books&v=glance}.

\bibitem{ting2007}
Jo-Anne Ting, Evangelos Theodorou, and Stefan Schaal.
\newblock Learning an outlier-robust kalman filter.
\newblock In Joost~N. Kok, Jacek Koronacki, Raomon Lopez~de Mantaras, Stan Matwin, Dunja Mladeni{\v{c}}, and Andrzej Skowron, editors, {\em Machine Learning: ECML 2007}, pages 748--756, Berlin, Heidelberg, 2007. Springer Berlin Heidelberg.

\bibitem{wiener1948cybernetics}
Norbert Wiener.
\newblock {\em Cybernetics: or Control and Communication in the Animal and the Machine}.
\newblock MIT Press, Cambridge, MA, 2 edition, 1948.

\bibitem{sokDex}
Jiahua Xu, Krzysztof Paruch, Simon Cousaert, and Yebo Feng.
\newblock Sok: Decentralized exchanges (dex) with automated market maker (amm) protocols.
\newblock {\em ACM Comput. Surv.}, 55(11), feb 2023.
\newblock \href {https://doi.org/10.1145/3570639} {\path{doi:10.1145/3570639}}.

\bibitem{oraclesProblem}
Adelyn Zhou.
\newblock Flash loans aren’t the problem, centralized price oracles are.
\newblock \url{https://www.coindesk.com/tech/2020/11/11/flash-loans-arent-the-problem-centralized-price-oracles-are/}.
\newblock Accessed: 2023-09.

\bibitem{zhou1996robust}
K.~Zhou, J.C. Doyle, and K.~Glover.
\newblock {\em Robust and Optimal Control}.
\newblock Feher/Prentice Hall Digital and. Prentice Hall, 1996.
\newblock URL: \url{https://books.google.com/books?id=RPSOQgAACAAJ}.

\bibitem{cartea23predictable}
Álvaro Cartea, Fayçal Drissi, and Marcello~Monga and.
\newblock Predictable losses of liquidity provision in constant function markets and concentrated liquidity markets.
\newblock {\em Applied Mathematical Finance}, 30(2):69--93, 2023.
\newblock \href {https://arxiv.org/abs/https://doi.org/10.1080/1350486X.2023.2277957} {\path{arXiv:https://doi.org/10.1080/1350486X.2023.2277957}}, \href {https://doi.org/10.1080/1350486X.2023.2277957} {\path{doi:10.1080/1350486X.2023.2277957}}.

\end{thebibliography}

\appendix
\section{Proof of \prettyref{thm:2}}

\begin{proof}
Note that, by definition, $\beta_t(p) = E[p_{ext}^t|\mathcal{H}_{t-1}, p^t_{trad}=p]$. If we further define a sequence of orthogonal random variables $e_t = p_{trad}^t - E[p_{trad}^t|\mathcal{H}_{t-1}]$, then we have
\begin{align}
    E[p_{trad}^t|\mathcal{H}_{t-1}] &= E[p_{ext}^{t}|\mathcal{H}_{t-1}] + E[\Delta p_{trad}^{t}|\mathcal{H}_{t-1}] = E[p_{ext}^{t}|\mathcal{H}_{t-1}],
\end{align}
by the independence of trader noise. This gives us that $e_t = p_{trad}^t - E[p_{ext}^t|\mathcal{H}_{t-1}]$. By orthogonality of the sequence $\langle e_\tau\rangle_{\tau=1}^t$, we have
\begin{align}
    E[p_{ext}^{t+1}|\mathcal{H}_{t}] &= \sum_{\tau=1}^t \frac{E[p_{ext}^{t+1} e_{\tau}]}{E[e_{\tau}^2]} e_\tau\\
    &= E[p_{ext}^{t+1}|\mathcal{H}_{t-1}] + \frac{E[p_{ext}^{t+1} e_{t}]}{E[e_{t}^2]} e_t\\
    &= E[p_{ext}^{t}|\mathcal{H}_{t-1}] + \frac{E[p_{ext}^{t+1} e_{t}]}{E[e_{t}^2]} (p_{trad}^t - E[p_{ext}^t|\mathcal{H}_{t-1}])\\
    &= (1-\frac{E[p_{ext}^{t+1} e_{t}]}{E[e_{t}^2]})E[p_{ext}^{t}|\mathcal{H}_{t-1}] + \frac{E[p_{ext}^{t+1} e_{t}]}{E[e_{t}^2]} p_{trad}^t.\label{eq:kalman_mod}
\end{align}
Here, we know that $\frac{E[p_{ext}^{t+1} e_{t}]}{E[e_{t}^2]} := K_t$ is the Kalman gain at time $t$, that can be computed iteratively using the Kalman filtering algorithm \cite{kalman}. Using this, the above equation \prettyref{eq:kalman_mod} can be equivalently written as
\begin{align}
    E[p_{ext}^{t}|\mathcal{H}_{t-1}, p^t_{trad}=p] &= (1-K_t)E[p_{ext}^{t}|\mathcal{H}_{t-1}] + K_t p.
\end{align}
Thus, we can solve the fixed point condition \prettyref{eq:main2} for the initial operating point as 
\begin{align}
    p &= \beta_t(p)\\
    p &= (1-K_t)E[p_{ext}^{t}|\mathcal{H}_{t-1}] + K_t p\\
    p &= E[p_{ext}^{t}|\mathcal{H}_{t-1}],
\end{align}
which gives us a unique solution to the initial operating point $p_0^t$ of the market. Thus, we can now write $\beta_t(p) = (1-K_t)p_0^t + K_t p$. 

Substituting this in the differential equation \prettyref{eq:main1} gives us
\begin{align}
    g_t'(p)(1-K_t)(p_0^t-p) + g_t(p) K_t - x_0^t K_t = 0
\end{align}
which gives us the general solution
\begin{align}
    g_t(p) = x_0^t - C_t (p-p_0^t)^{\frac{K_t}{1-K_t}},
\end{align}
where the constant $C_t>0$ is a solution to the boundary condition $y_0^t = -\int_0^{p_0^t} pg_t'(p)dp$. While this function is well defined for $p\geq p_0^t$, it is not well defined for $p<p_0^t$. However, if we set $C_t = 0$ by forcing $g_t(p)$ to be constant on either side of $p=p_0^t$, then we get the solution in the statement of the theorem.

\end{proof}

\section{Proof of \prettyref{thm:3}}

\begin{proof}
    As before, we derive an expression for $\beta_t(p)$ first. In this case, we have
    \begin{align}
        \beta_t(p) &= E[p_{ext}^t|\mathcal{H}_{t-1}, p_{trad}^t=p]\\
        &= E[e^{\log p_{ext}^t}|\mathcal{H}_{t-1}, p_{trad}^t=p]
    \end{align}
    We know that $E[\log p_{ext}^t|\mathcal{H}_{t-1}, p_{trad}^t=p] = (1-K_t)E[\log p_{ext}^t|\mathcal{H}_{t-1}] + K_t \log p$, where $K_t$ is the Kalman gain corresponding to the process $\langle \log p_{trad}^\tau \rangle_{\tau=1}^t$. 
Using the formula for change of variables of random variables, we get
\begin{align}
    E[e^{\log p_{ext}^t}|\mathcal{H}_{t-1}, p_{trad}^t=p]
    &=\exp\left(E[{\log p_{ext}^t}|\mathcal{H}_{t-1}, p_{trad}^t=p] + \frac{P_{t|t}}{2}\right)\\
    &= \exp\left((1-K_t)E[\log p_{ext}^t|\mathcal{H}_{t-1}] + K_t \log p + \frac{P_{t|t}}{2}\right).
\end{align}
where $P_{t|t}$ is the variance of the estimate $E[\log p_{ext}^t|\mathcal{H}_{t-1}]$, and $K_t$ is the Kalman gain corresponding to the same estimate.
    We can now use the fixed point condition \prettyref{eq:main2} and get
    \begin{align}
        p = \exp\left((1-K_t)E[\log p_{ext}^t|\mathcal{H}_{t-1}] + K_t \log p + \frac{P_{t|t}}{2}\right), 
    \end{align}
    which, when solved further, gives us the initial operating point of the curve as 
    \begin{align}
        p_0^t &= \exp{\left(E[\log p_{ext}^t|\mathcal{H}_{t-1}] + \frac{P_{t|t}}{2(1-K_t)}\right)}.
    \end{align}
    Defining $\kappa_t = \exp{\left(\frac{P_{t|t}}{2(1-K_t)}\right)} $, this leads to the following differential equation for the curve -
    \begin{align}
       ( \kappa_t p^{K_t} - p) g_t'(p) + \kappa_t K_t p^{K_t-1}(g_t(p) - x_0^t) = 0
    \end{align}
    which has the solution
    \begin{align}
        g_t(p) = x_0^t - C_t (p^{1-K_t} - \kappa_t)^{\frac{K_t}{1-K_t}}p^{K_t}
    \end{align}
The rest of the proof follows in a way similar to \prettyref{thm:2}.

\end{proof}

\section{Proof of \prettyref{thm:static_comparison}}

\begin{proof}

    For a static curve, $p_{SC}^T = p_{trad}^T$, which gives us the MSE as $\eta^2$ by definition. 
    
    For Algorithm \prettyref{alg:kf}, we see that the variance of the estimate is $P_{0|0} = \sigma^2$. After that, we update the variance according to Line \ref{line1} in the algorithm. After trade $t>1$, the Kalman gain is $K_t = \frac{P_{t-1|t-1}}{P_{t-1|t-1} + \eta^2}$, which implies that $1-K_t = \frac{\eta^2}{P_{t-1|t-1} + \eta^2}$. This gives us 
    \begin{align}
        P_{t|t} &= \frac{\eta^2}{P_{t-1|t-1} + \eta^2} P_{t-1|t-1},
    \end{align}
    or, equivalently
    \begin{align}
        \frac{1}{P_{t|t}} &= \frac{1}{\eta^2} + \frac{1}{P_{t-1|t-1}}.
    \end{align}
    Writing the sum for $t=T$ till $t=0$ gives us
    \begin{align}
        \frac{1}{P_{t|t}} &= \frac{T}{\eta^2} + \frac{1}{\sigma^2},
    \end{align}
    which proves the desired result.
\end{proof}

\section{Proof of \prettyref{thm:static_implied}}

\begin{proof}
    We can adapt the Kalman filter updates and apply them to Equations \prettyref{eq:5update1} and \prettyref{eq:5update2}. Since $p_{trad}^{t-1}$ is observed exactly by the market maker, we have $P_{t-1|t-1} = 0$. This gives us the Kalman gain as
    \begin{align}
        K_t &= \frac{P_{t-1|t-1}+\sigma^2}{P_{t-1|t-1}+\sigma^2+\eta^2}\\
        &= \frac{0+\sigma^2}{0+\sigma^2+\sigma^2(1/\theta - 1 )}\\
        &= \theta
    \end{align}
    The Kalman estimate then becomes
    \begin{align}
        E[\log p_{ext}^t |\mathcal{H}_{t-1}, p_{trad}^t=p] &= (1-\theta) \log p_{trad}^{t-1} + \theta \log p
    \end{align}
    This implies that we can write $\beta_t(p)$ as
    \begin{align}
        \beta_t(p) &= E[ p_{ext}^t |\mathcal{H}_{t-1}, p_{trad}^t=p] \\
        &= \exp\left((1-\theta) \log p_{trad}^{t-1} + \theta \log p + \frac{\sigma^2}{2(1-\theta)} \right)
    \end{align}
    Note that, for static curves, the previous trader price is directly used as the initial operating point for the next trader. This means that $p_0^t=p_{trad}^{t-1}$. Substituting this gives us
    \begin{align}
        \beta_t(p) &= p^\theta (p_0^t)^{1-\theta} e^{\frac{\sigma^2}{2(1-\theta)}}\label{eq:belief2}
    \end{align}
    We now compare \prettyref{eq:belief1} and \prettyref{eq:belief2} for $\sigma\ll 1$. By \prettyref{eq:5update1}, we know that as $\sigma \rightarrow 0$, $\frac{p_0^t}{p} \rightarrow 1$. Using this in the last term in \prettyref{eq:belief1} gives us
    \begin{align}
        \lim_{p_0^t/p\rightarrow 1} \frac{1-(p_0^t/p)^\theta}{1 - (p_0^t/p)^{1-\theta}} &= \frac{\theta}{1-\theta}
    \end{align}
    which gives us $\beta_t(p) = p^\theta (p_0^t)^{1-\theta}$ in \prettyref{eq:belief1}. The same expression is also obtained from \prettyref{eq:belief2} as $\sigma\rightarrow 0$.
\end{proof}



\end{document}